\documentclass[12pt]{iopart}
\usepackage{lineno}
\usepackage{graphicx}% Include figure files
\usepackage{graphics}% Include figure files
\usepackage{hyperref}
\usepackage{color}
\usepackage{times}

\begin{document}

\title[]{%On the frequency of hydrodynamic perturbations. From the early transient through the intermediate term to the asymptotic state.//
New observations in  transient hydrodynamic perturbations. Frequency jumps, intermediate term and spot formation.}
\author{Stefania Scarsoglio, Francesca De Santi and Daniela Tordella\footnote{Corresponding
author: daniela.tordella@polito.it}}

\address{Department of Mechanical and Aerospace Engineering, Politecnico di Torino, Italy}
%\ead{custserv@iop.org}

\begin{abstract}
%We present recent findings concerning the frequency in the transient evolution of three-dimensional perturbations in sheared flows.
    Sheared incompressible flows are usually considered non-dispersive media. As a consequence, the frequency evolution in transients has received much less attention than the wave energy density or growth factor. By carrying out a large number of long term transient simulations for longitudinal, oblique and orthogonal perturbations in the plane channel and wake flows, we could observe a complex time evolution of the frequency. The complexity is mainly associated to jumps  which appear quite far along within the transients. We interpret these jumps as the transition between the early transient and the beginning of an  intermediate term that reveals itself for times large enough for the influence of the fine details of the initial condition to disappear. The intermediate term leads to the asymptotic exponential state and
    %In the presence of a wall, regardless of the symmetry of the initial condition, the frequency of non orthogonal waves jumps to values that can be $30-40\%$ higher than those observed in the early transient. In the case of the \textcolor{red}{plane} wake, which is the free flow considered in this paper, the situation is similar, but the jumps are generally lower.
    %Jumps are often accompanied by oscillations that begin in the early transient, may last throughout the intermediate term and disappear when the asymptotic exponential state is reached.
has a duration one order of magnitude longer than the early term, which indicates the existence of an intermediate asymptotics. Since after the intermediate term perturbations die out or blowup, the mid-term can be considered as the most probable state in the perturbation life.

    Long structured transients can be related to the spot patterns commonly observed  in subcritical transitional wall flows.  By considering a large group of three-dimensional waves in a narrow range of wave-numbers, we superposed them in a finite temporal window with oblique and longitudinal waves randomly delayed with  respect to an orthogonal wave which is supposed to sustain the spot formation with its intense transient growth. We show that in this way it is possible to recover the linear initial evolution of the pattern characterized by the presence of longitudinal streaks. We also determined the asymptotic frequency and phase speed  wave-number spectra for the channel flow at Reynolds numbers 500 and 10000 and for the wake at Reynolds numbers 50 and 100. In both cases, for long travelling oblique and longitudinal waves a narrow dispersive range can be observed. This mild dispersion can in part explain the different propagation speeds of the backward and forward fronts of the spot.

\end{abstract}

\pacs{*43.20.Ks, 46.40.Cd, 47.20.-k, 47.20.Gv, 47.35.De, 84.40.Fe, 47.54.-r, 47.20.Ft}
\submitto{\NJP}
\maketitle

\tableofcontents

\linenumbers
\section{Introduction}

Although both Kelvin \cite{K1887a,K1887b} and Orr \cite{O1907a,O1907b} recognized that the early transient contains important information, only in recent decades have many contributions been devoted to the study of the transient dynamics of three-dimensional perturbations in shear flows \cite{SH2001,CJJ2003}. For a long time linear modal analysis, developed by Orr \cite{O1907a,O1907b} and Sommerfeld \cite{S1908}, has been considered a sufficient and efficient tool to analyze hydrodynamic stability, see for instance the historical paper by Taylor (1923) \cite{T1923}. More recently it has been observed that the early stages of the disturbance evolution can deeply affect the stability of the flow \cite{BF1992,Bergstrom2005,G1991,CD90,RH1993}. In fact, early algebraic growth can show exceptionally large amplitudes long before an exponential mode is able to set in. It is believed that this kind of behaviour is able to promote rapid transition to fluid turbulence, a phenomenon known as bypass transition \cite{BB2009,HLJ1993,LJJC1999,L1996}. Transient decay of asymptotically unstable waves is also possible, which makes the situation rich and complex at the same time. An example of this possible scenario is represented by pipe flow. Linear modal analysis assures stability for all the Reynolds numbers \cite{DR2002}, but this result is in contrast with the experimental evidence, which shows that the flow becomes turbulent at sufficiently large Reynolds numbers. The disagreement between the linear modal prediction and laboratory results has motivated several recent works \cite{FE2003,Hof2004,DBL2010} that focus on transient travelling waves and their link to the transition process. In general, it is now considered possible that inside the transient life of travelling waves some important events for the stability of the flow can take place.

In this work, we focus on the temporal evolution of the wave frequency in two archetypical shear flows, the plane channel flow and the bluff-body plane wake. Probably due to the fact that incompressible shear flows are viewed as non-dispersive media,
the frequency transient has been poorly investigated so far. For instance, in the wake flow the attention was mainly devoted to the frequency of vortex shedding for the most unstable spatial scales \cite{W89,SS1990}. Only very recently, subcritical wake regimes (up to values $30\%$ below the critical value) of the vortex shedding of transiently amplified perturbations were studied by considering the spatio-temporal evolution of wave packets \cite{M2011}.

The situation is quite different within the context of atmosphere and climate dynamics. Here, the interaction between low-frequency and high-frequency phenomena, which is related to the existence of very different spatial and temporal scales, is believed to be one of the main reasons for planetary-scale instabilities \cite{S2002,Na1997}. However, due to the inherent strong nonlinearity, the evolution of single scales cannot be observed in the geophysical systems and thus also these studies usually do not account for the frequency transient evolution of a single wave.

Perturbation transient lives show trends which are not easily predictable. Intense  transient growth as well as transient decays  are just some relevant examples of the observed scenery. In this study, the transition between the early transient and the asymptotic state is observed in detail. It was then possible to highlight an unexpected phenomenon: this transition does  not occur smoothly. Frequency jumps appear at an intermediate stage located in between the beginning of the time evolution and the setting of the asymptotic state. We interpret the appearance of the frequency jumps, which are usually preceded and followed by modulating fluctuations, as the beginning of the dynamic process yielding the final state. In so doing we introduce an intermediate term, which separates the early and final stages of the evolution. We observe that the length of the early term is much shorter than that of the intermediate term.

Spot formation is one of the more commonly observed phenomenology related to the existence of long transients. Here, we try  to relate spot patterns to the combination of two facts: - the very long transient of  orthogonal perturbations and their intense transient growth or, for low enough Reynolds numbers, their least monotonic decay, - the necessary presence of oblique waves with wavenumber close to that of the orthogonal perturbation. Spots has been studied in different wall-bounded shear flows by means of laboratory and numerical experiments. Because of its simplicity, plane Couette flow is perhaps the most studied case to understand pattern formation, that is the coexistence in space and time of laminar and transitional regions \cite{DHB1992,BT2005,Prigent2002}. Probably due to an overvaluation of the results associated to the modal theory  (this flow is asymptotically  stable for all the Reynolds numbers), for long the transition to turbulence of the Couette flow was considered a puzzle. When the algebraic transient growth importance was recognized as a possible promoter of the nonlinear coupling, a critical Reynolds number for the Couette flow, $Re_c \sim 325$, has been individuated. Below this value, spots decay in the long-time, while above this threshold the turbulent motion is sustained \cite{Bottin1998, DSH2010}.

When analyzing the shape and the structure of the spot, Lundbladh and Johansson \cite{LJ1991} classify the Couette spot as an intermediate case between Poiseuille and boundary layer spots. Dauchot and Daviaud \cite{DD1995} find the spot evolution for the Couette flow very similar to that observed for the plane Poiseuille spot. Measurements on the propagation velocity of the spot \cite{Tillmark,HA1987} show a quite high level of variability inside the spot, ranging from 10\% up to 60\% of the reference velocity of the flow \cite{CWP1982,CCD1978,AHA1986}. In general, the rear part moves slower than the front part.

In this work, we exploit the collection of three-dimensional perturbations computed both in the plane channel and wake flows to reproduce the formation of spots by the linear superposition of a large number of waves around an observation point. Conceptually, we want to simulate the transient triggering of a fan of oblique waves (in the $[-\pi/2, \pi/2]$ range of angles about the mean flow direction) by means of an orthogonal standing wave. The superposition is randomized by delaying the wave entrance in a temporal window lasting order 10 physical time scales.
The tight parametrization on the wavenumber and obliquity angle carried out to explore the transient is then exploited to get information on the asymptotic behaviour of the phase velocity.

The organization of the paper is as follows. To observe the life of a perturbation, an initial-value problem is formulated, see Section 2 and the Appendix A.  The frequency behaviour in the transient is then described in Section 3, relevant supplemental details are in the Appendix B. In section 4, a discussion on the existence of the an intermediate term and its empirical determination is presented: this is accompanied by discussion on the role played by orthogonal waves and the spot formation process. Section 5 deals with asymptotic properties: wavenumber spectra and phase speed obliquity dependence. Conclusion remarques follow in section 6.

\section{Background}

\begin{figure}
\begin{center}
\begin{minipage}[b]{0.72\columnwidth}%\begin{tabular}{c c}
\hspace{-1.3cm}
\includegraphics[width=\columnwidth]{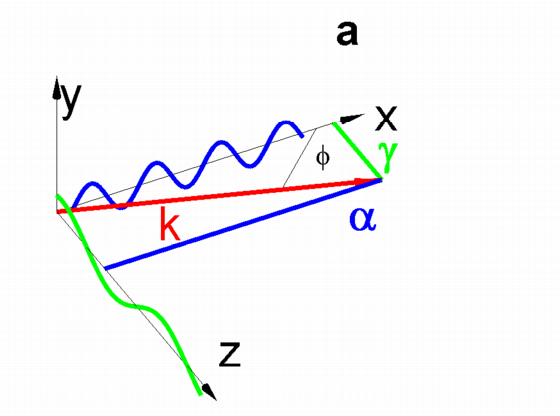}
\end{minipage}
\hspace{-3.5cm}
\begin{minipage}[b]{0.28\columnwidth}
\includegraphics[width=1.2\columnwidth]{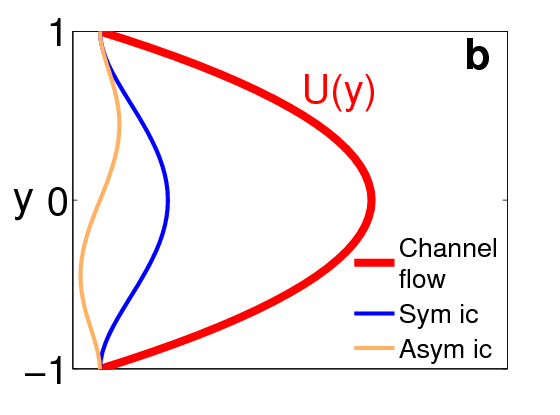}\\
\includegraphics[width=1.2\columnwidth]{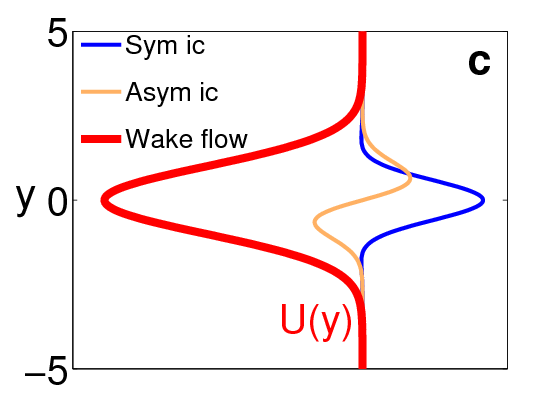}
\end{minipage}
\end{center}
\hspace{-1.3cm}
%\vskip -6mm
\caption {(a) Perturbation geometry scheme. Perturbations propagate in the direction of
the polar wavenumber, $k =\sqrt{\alpha^2+\gamma^2}$, ($\alpha$ and $\gamma$ are the streamwise (x) and spanwise (z)
wavenumbers, respectively). $\phi$ is the angle of obliquity with respect to the basic flow. (b)-(c) Base flow velocity profiles, $U(y)$ (thick curves), and symmetric and antisymmetric initial conditions of the perturbation transversal velocity, the $\hat{v}$ component along the $y$ direction, $\hat{v}(y, t = 0)$ (thin curves).}
\label{Fig_1}
\end{figure}

The transient and longterm behaviour is studied using the initial-value problem formulation. We consider two typical shear flows, the plane Poiseuille flow, the archetype of wall flows, and the plane bluff-body wake, one of the few free flow archetypes (see Fig. \ref{Fig_1}b and \ref{Fig_1}c, respectively).
The viscous perturbation equations are combined in terms of the vorticity and velocity \cite{CD90}, and then solved by means of a combined Fourier--Fourier (channel) and Laplace--Fourier (wake) transform in the plane normal to the basic flow. This slightly different formulation is due to the fact that the channel flow is homogeneous in the streamwise ($x$) and spanwise ($z$) directions, while the wake flow is homogeneous in the $z$ direction and slightly inhomogeneous in the $x$ direction. For the wake, the domain is defined for $x\geq0$ ($x=0$ is the position of the body which generates the flow).

\subsection{Initial conditions}
Unlike traditional methods where travelling wave normal modes
are assumed as solutions, we follow \cite{CJLJ97} and use arbitrary initial conditions that can
be specified without having to recur to eigenfunction expansions. Within our framework, for any initial small-amplitude three-dimensional disturbance, this method allows the determination of the complete temporal behaviour, including both the early and intermediate transients and the long-time asymptotics. It should be recalled that an arbitrary initial disturbance could be expanded in terms of the complete set of discrete and continuum eigenfunctions, as it was demonstrated in the more general case of open flows
by Salwen and Grosch \cite{SG1981}. In bounded flows, in fact, it would be sufficient to expand in terms of discrete  eigenfunctions.

In literature, various initial conditions were used to explore transient behaviour at subcritical
Reynolds numbers. The important physical issue is however the ability to make, in a
simple manner, arbitrary specifications. Since a normal mode decomposition provides
a complete set of eigenfunctions, it is true that any arbitrary specification can
(theoretically) be written in terms of an eigenfunction expansion. Nevertheless, it should be noted that there is
nothing special about the eigenfunctions when it comes to specifying initial conditions. They simply represent the most convenient means of specifying the long-term solution.

Furthermore, the adoption of non-orthogonal eigenfunctions in the try to build
any real arbitrary initial condition introduces unnecessary mathematical complications.
Physically, it  seems that the natural issues affecting the initial specification are whether the disturbances  are, first, symmetric or antisymmetric and, second, local or more distributes across the basic profile of the flow. The cases we used here satisfy both of these needs and use functions that can be employed to represent any arbitrary initial distribution. If not otherwise specified, with symmetric and antisymmetric conditions we intend the initial conditions as specified in the A Appendix and shown in Fig. 1b-c.

\subsection{Perturbative analysis}
The exploratory analysis is carried out with respect to physical quantities, such as the polar wavenumber, $k$, the angle of obliquity with respect to the basic flow plane, $\phi$, the symmetry of the perturbation with respect to $y$ (which is the coordinate orthogonal to the wavenumber vector), and the flow control parameter, $Re$. For the channel flow, $Re = U_0 h/\nu$ is the Reynolds number, where $h$ is the channel half-width, $U_0$ the centreline velocity and $\nu$ the kinematic viscosity. The Reynolds number for the plane wake is defined as $Re = U_f D/\nu$, where $D$ is the body diameter, $U_f$ is the free stream velocity and $\nu$ the kinematic viscosity. We define the longitudinal wavenumber, $\alpha=k \cos(\phi)$, and the transversal wavenumber, $\gamma= k \sin (\phi)$, see Fig. \ref{Fig_1}a. The perturbation and the flow schemes are presented in Fig. \ref{Fig_1} (a,b,c). More details on the formulation are provided in the Appendix A. The basic eddy turn over time is defined as $ h/U_0$ and $D/U_f$ for the channel and wake flows, respectively.

\noindent As longitudinal observation points we selected for the plane wake, which is near parallel, two positions downstream of the body: $x_0=10$, which is a position inside the intermediate part of the wake spatial development, and $x_0=50$, which is a location inside the far field. The frequency has been evaluated in these sections at a transversal position, which in the following is called $y_0$. For the channel flow, which is homogeneous in the streamwise direction, it is sufficient to specify the transversal position, $y_0$.

\begin{figure}
\begin{center}
\begin{minipage}[]{0.32\columnwidth}	
\includegraphics[width=\columnwidth]{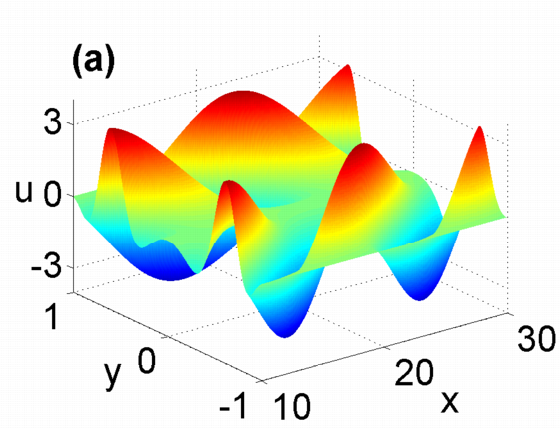}
\end{minipage}
\begin{minipage}[]{0.32\columnwidth}	
\includegraphics[width=\columnwidth]{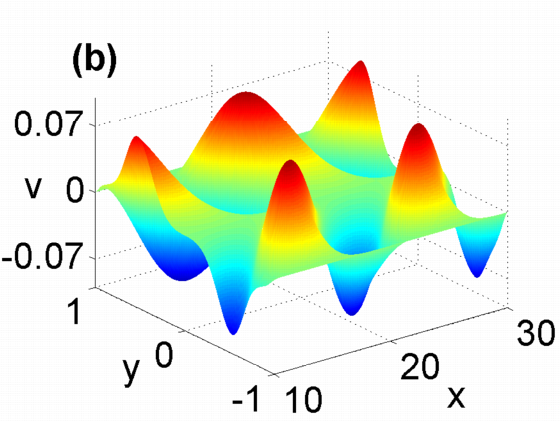}
\end{minipage}
\begin{minipage}[]{0.32\columnwidth}	
\includegraphics[width=\columnwidth]{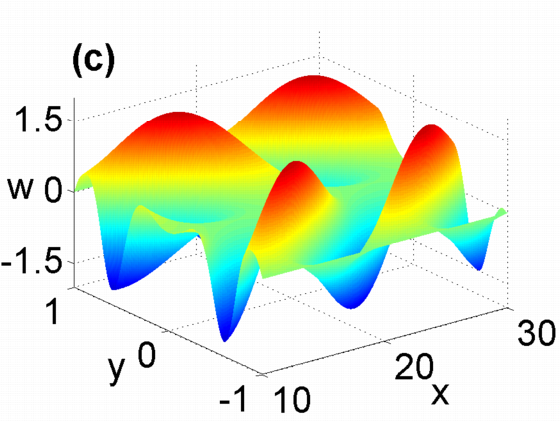}
\end{minipage}
\begin{minipage}[]{0.32\columnwidth}	
\includegraphics[width=\columnwidth]{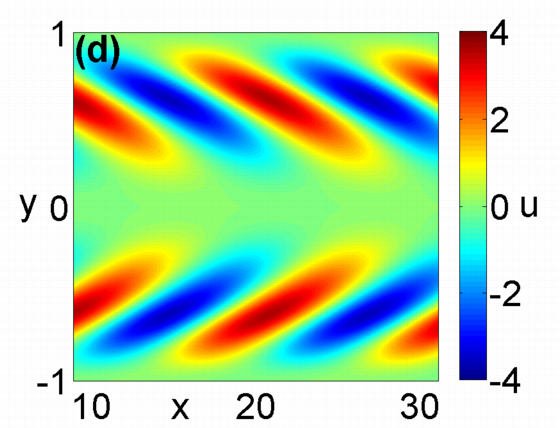}
\end{minipage}
\begin{minipage}[]{0.32\columnwidth}	
\includegraphics[width=\columnwidth]{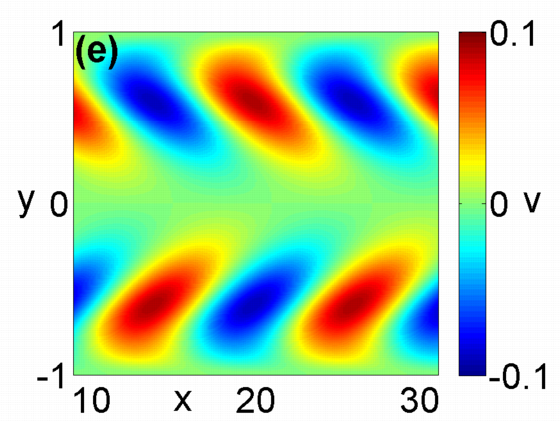}
\end{minipage}
\begin{minipage}[]{0.32\columnwidth}	
\includegraphics[width=\columnwidth]{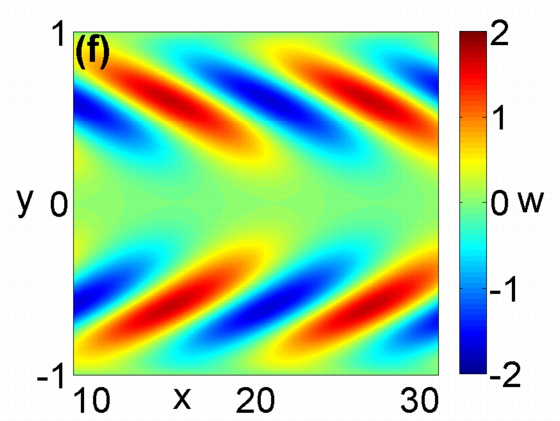}
\end{minipage}
\end{center}
%\vskip -6mm
\caption {Channel flow. Perspective (panels a-b-c panels) and projection (panels d-e-f) view visualizations of the perturbation velocity field in the physical plane $(x, y)$, $Re=10000$, $t=20$, angle of obliquity $\phi=3/8 \ \pi$, antisymmetric initial condition, $k=1.5$. Panels (a)-(d) refer to the velocity component $u$, panels (b)-(e) to $v$, and panels (c)-(f) to $w$. The wavelength is normalized over the flow external spatial scale, the channel half-height $h$. The time is normalized over the flow relevant eddy turnover time, $h/ U_0$, where $U_0$ is the centerline velocity.}
\label{Fig_2}
\end{figure}

\begin{figure}
\begin{center}
\begin{minipage}[]{0.32\columnwidth}	
\includegraphics[width=\columnwidth]{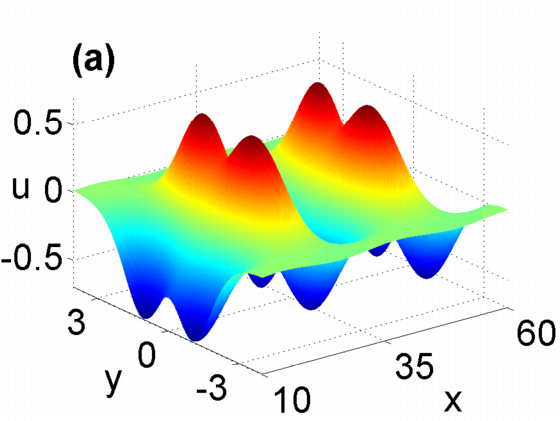}
\end{minipage}
\begin{minipage}[]{0.32\columnwidth}	
\includegraphics[width=\columnwidth]{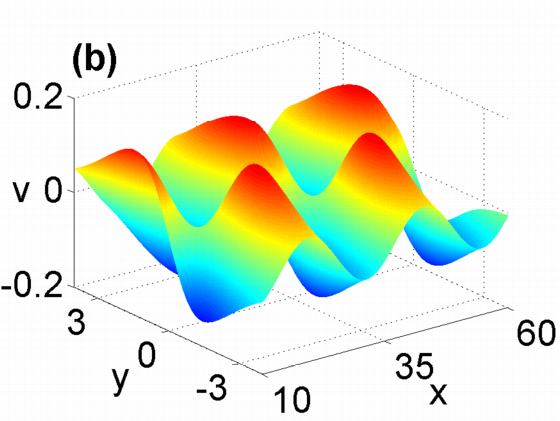}
\end{minipage}
\begin{minipage}[]{0.32\columnwidth}	
\includegraphics[width=\columnwidth]{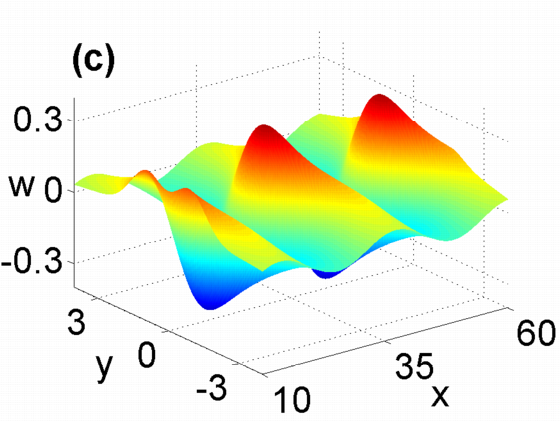}
\end{minipage}
\begin{minipage}[]{0.32\columnwidth}	
\includegraphics[width=\columnwidth]{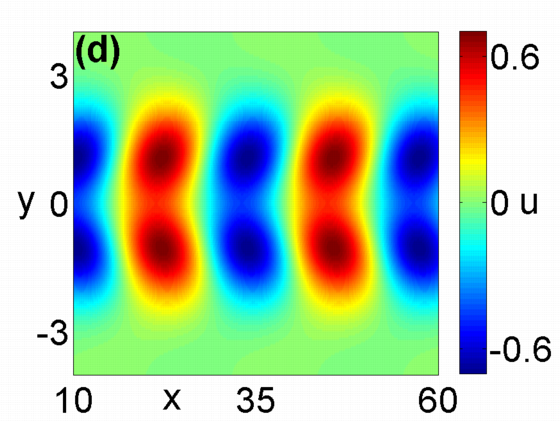}
\end{minipage}
\begin{minipage}[]{0.32\columnwidth}	
\includegraphics[width=\columnwidth]{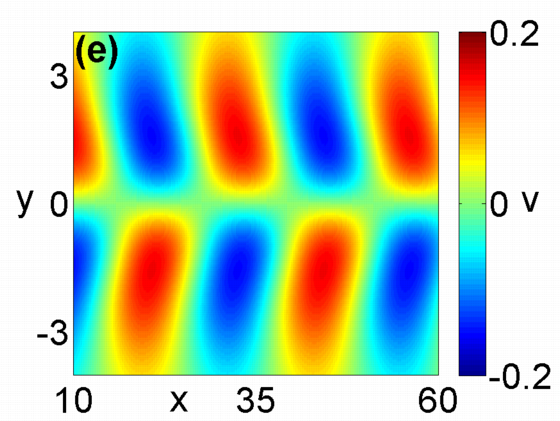}
\end{minipage}
\begin{minipage}[]{0.32\columnwidth}	
\includegraphics[width=\columnwidth]{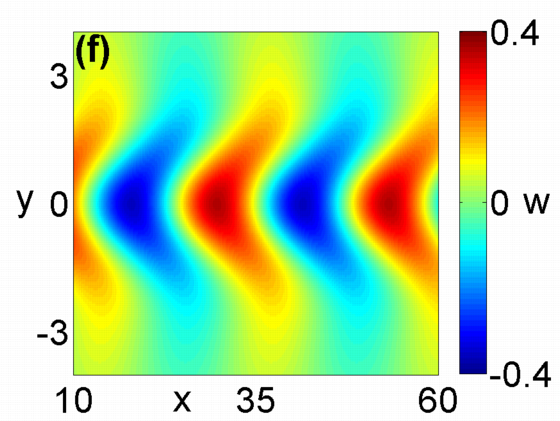}
\end{minipage}
\end{center}
%\vskip -6mm
\caption {Wake flow. Perspective (panels a-b-c panels) and projection (panels d-e-f) view visualizations of the perturbation velocity field in the physical plane $(x, y)$, $Re=100$, wake section: 50 body lengths downstream, $t=45$, angle of obliquity $\phi=3/8 \ \pi$, antisymmetric initial condition, $k=0.7$. Panels (a)-(d) refer to the velocity component $u$, panels (b)-(e) to $v$, and panels (c)-(f) to $w$. The wavelength is normalized over the flow external spatial scale, the bluff body diameter $D$. The time is normalized over the flow relevant eddy turnover time, $D/U_f$, where $U_f$ is the free stream velocity.}
\label{Fig_3}
\end{figure}

\noindent In Figures \ref{Fig_2} and \ref{Fig_3}, one can see two examples of  visualizations (a perspective  and a projection view) of the perturbation velocity field, ($u, v, w$), for the channel and the plane wake flows, respectively. The visualizations are displayed in the physical space $x, y$ ($u, v, w$ are obtained by a discrete 2D anti-transform of the solved quantities, $\hat{u}, \hat{v}, \hat{w}$), for an oblique perturbation with wavenumber equal to 1.5 for the channel and equal to 0.7 for the wake flow. The wave lengths are normalized over the channel half height and the body diameter, respectively. The time, $t$, which appears in the figures is the independent temporal variable normalized with respect to the basic flow eddy turn over time.

To measure the growth of the perturbations, we define the kinetic energy density,

\begin{equation}
e(t; \alpha, \gamma) = \frac{1}{4 y_f} \int_{-y_f}^{+y_f}
(|\hat{u}|^2
+ |\hat{v}|^2 + |\hat{w}|^2) dy,
\end{equation}

\noindent where $-y_f$ and $y_f$ are the computational limits of the domain, while $\hat{u}(y, t; \alpha, \gamma)$, $\hat{v}(y, t; \alpha, \gamma)$ and $\hat{w}(y, t; \alpha, \gamma)$ are the transformed velocity components of the perturbed field. For the channel flow, which is bounded, the computational limits coincide with the walls ($y_f=1$). The wake is an unbounded flow and the value $y_f$ is defined so that the numerical solutions
are insensitive to further extensions of the computational domain size ($y_f = 20$ for waves with $k>1$ and $y_f = 40$ for longer waves). We then introduce the amplification factor, $G$, as the kinetic energy density normalized with respect to its initial value,

\begin{equation}
G(t; \alpha, \gamma) = e(t; \alpha, \gamma)/e(t=0; \alpha, \gamma).
\end{equation}

\noindent Assuming that the temporal asymptotic behaviour of the linear perturbations is exponential, the temporal growth rate, $r$, can be defined \cite{CJJ2003} as

\begin{equation}
r(t; \alpha, \gamma) = log(e)/(2t).
\label{tgr}
\end{equation}

\noindent The frequency, $\omega$, of the perturbation is
defined as the temporal derivative of the unwrapped wave phase, $\theta(y, t; \alpha, \gamma)$, at a specific spatial point along the $y$ direction. The wrapped phase,

\begin{equation}
\theta_w(y, t; \alpha, \gamma) = arg(\hat{v}(y, t; \alpha, \gamma)),
\end{equation}

\noindent is a discontinuous function of $t$ defined in $[-\pi,+\pi]$, while the unwrapped phase, $\theta$, is a continuous function obtained by introducing a sequence of $2 \pi$ shifts on the phase values in correspondence to the periodical discontinuities, see figures 4 and 5. In the case of the wake we use as reference transversal observation point $y = y_0 = 1$, and in the case of the channel flow the point $y = y_0 = 0.5$. The frequency \cite{STC09} is thus

\begin{equation}
\omega(t; y_0, \alpha, \gamma) = |d \theta(t; y_0, \alpha, \gamma)|/dt.
\end{equation}

\noindent It should be noted that when $r$ and $\omega$ become constant, the asymptotic state is reached.

The phase velocity is defined as

\begin{equation}
{\bf C}=(\omega/ k) \hat{\textbf{k}},
\end{equation}

\noindent where $\hat{\textbf{k}} = (\cos(\phi), \sin(\phi))$ is the unitary vector in the $k$ direction, and represents the rate at which the phase of the wave propagates in space.

\section{Frequency transient}

Transient dynamics offer a great variety of different behaviours and phenomena, which are not easy to predict \emph{a priori}. It is interesting
to note that these phenomena develop in the context of the linear dynamics, where interaction among different perturbations (and even self-interaction) is absent.

As an example, the findings about the angular frequency jumps below described  make the frequency transient complex, which means to make complex the time history of the phase speed for all the longitudinal and oblique waves. The orthogonal waves only stand apart  since they do not oscillate in time.  If one considers a swarm of perturbations distributed over a large range of wave-numbers and obliquity angles, since the frequency jumps for each wave arrive at different instants inside the transients, one may imagine that the overall evolution, their sum, will be exceedingly complex. Even as nearly chaotic, which of course is not at all true, since we are working in the linear context.
In any case,  the concept: one wavelength and propagation direction, one frequency, is oversimplified and  becomes applicable in asymptotic conditions only.

\begin{figure}
\vspace{-5mm}
\begin{minipage}[]{0.49\columnwidth}
\includegraphics[width=\columnwidth]{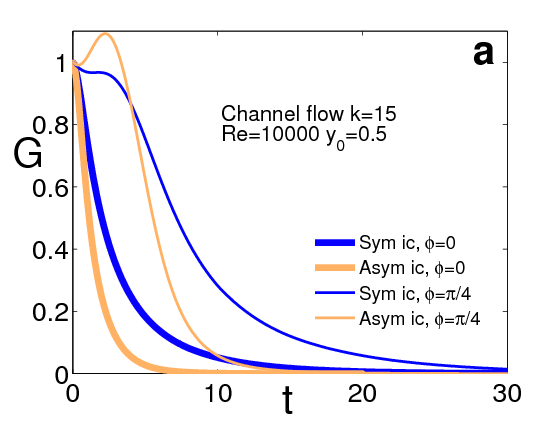}
\end{minipage}
\hspace{-.8cm}
\begin{minipage}[]{0.49\columnwidth}
\includegraphics[width=\columnwidth]{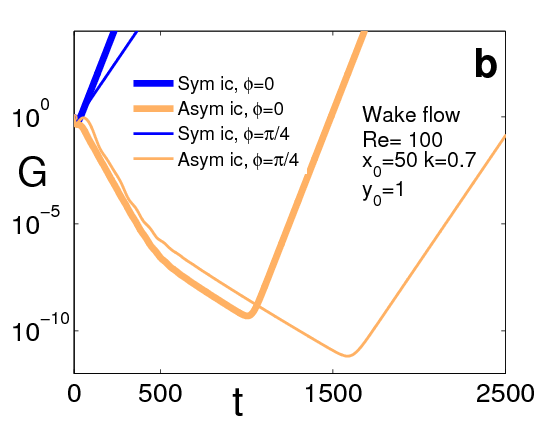}
\end{minipage}
\begin{minipage}[]{0.49\columnwidth}
\includegraphics[width=\columnwidth]{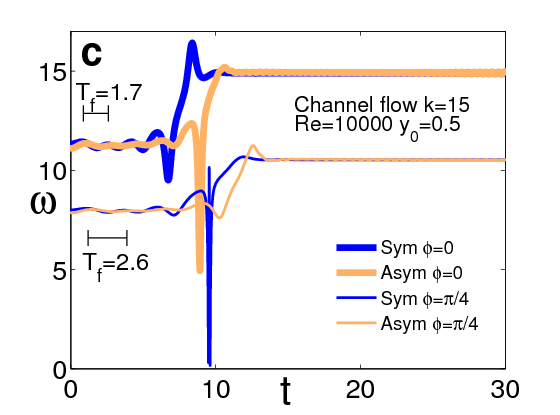}
\end{minipage}
\begin{minipage}[]{0.49\columnwidth}
\includegraphics[width=\columnwidth]{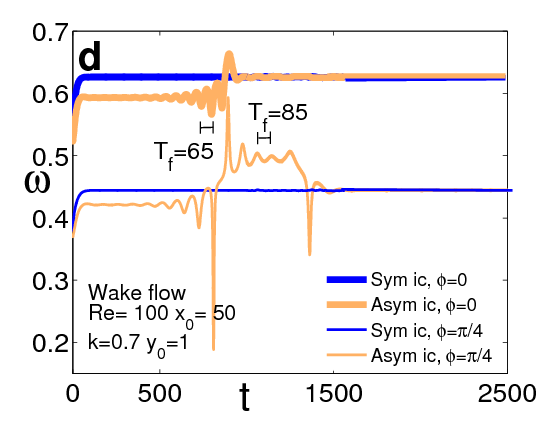}
\end{minipage}
\begin{minipage}[]{0.49\columnwidth}
\includegraphics[width=\columnwidth]{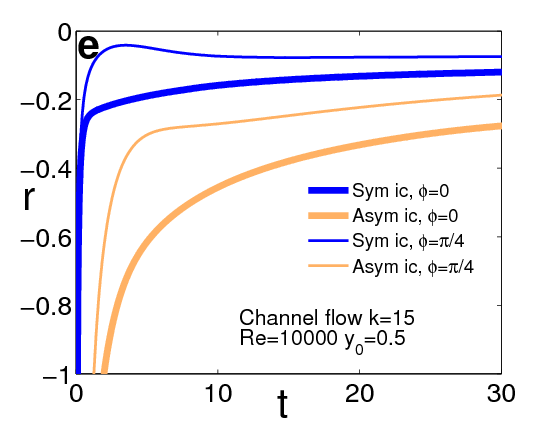}
\end{minipage}
\begin{minipage}[]{0.49\columnwidth}
\includegraphics[width=\columnwidth]{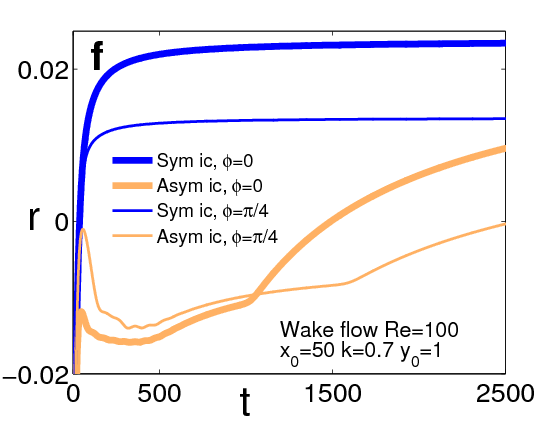}
\end{minipage}
\caption{Transient lives of the perturbations observed through the amplification factor, $G$ (top), the frequency, $\omega$ (middle), and the temporal growth rate, $r$ (bottom), at a fixed polar wavenumber and varying the obliquity ($\phi = 0, \pi/4$) and the symmetry of the initial conditions. Left column: channel flow, $k = 15$, $Re = 10000$. Right column: wake flow, $Re = 100$, $k = 0.7$: here the wake profile is observed at a distance past the body equal to 50 length scales, $x_0 = 50$. The angular frequency $\omega$ is computed at a distance from the wall equal to 1/4 of the channel width, ($y_0 = 0.5$, panel c), and at a distance equal to one body length, D, from the centre of the wake flow ($y_0 = 1$, panel d). The quantity $T_f$ (see panels c-d) indicates the temporal periodicity related to the frequency fluctuations observed in the early and intermediate dynamics.}
\label{transient}
\end{figure}

%\subsection{Frequency jumps}
We start the discussion by presenting an overview for transient dynamics, comparing evolutions of amplification factor, $G$, the frequency, $\omega$, and the temporal growth rate, $r$, see Fig. \ref{transient}.

\begin{figure}
\begin{center}
\includegraphics[width=0.7\columnwidth]{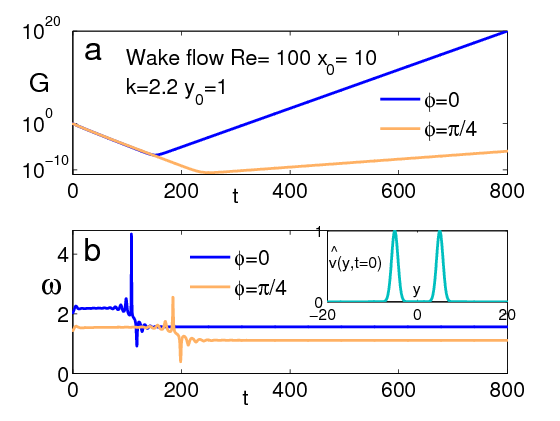}
\end{center}
	\caption{Frequency jumps observed for symmetric initial conditions in the wake flow ($Re=100$, position in the flow: $x_0=10$ and $y_0=1$, obliquity angles $\phi=0, \pi/4$, $k=2.2$): normalized energy (top panel) and frequency (bottom panel) transients. The inset in the bottom panel shows the specific symmetric initial condition in terms of $\hat{v}(y,t=0)=\exp(-(y-5)^2)+\exp(-(y+5)^2)$ considered here.}
	\label{sym_wake_frequency}
\end{figure}

The wavenumber is $k=15$ for the channel flow and $k=0.7$ for the wake flow. Two angles of obliquity ($\phi=0, \pi/4$), with symmetric and antisymmetric initial conditions, are considered. One can see that oblique stable waves present maxima of energy in time before being asymptotically damped (see in particular the case of channel flow in panel a). On the contrary, non-orthogonal perturbations can be significantly damped before an ultimate growth occurs (see the wake flow in panel b). An important observation is that quite far along within the transient, frequency discontinuously jumps to a value close to the asymptotic value ($\omega_a$), which is in general higher than the average value in the transient ($\omega_t$). The relative variation between transient and asymptotic values can change from a few percentages (about $5\%$) in case of the wake flow (see panel d), to values up to $30-40\%$ in the case of the channel flow (see panel c). In Fig. \ref{transient}d, frequency jumps for the wake are only observable for antisymmetric initial conditions. However, this is not true in general. Indeed, different symmetric inputs (see Fig. \ref{sym_wake_frequency}) can lead to discontinuous frequency transients as well. Moreover, we observe that even if symmetric and antisymmetric perturbations can have slightly different frequency values along the transient, they always reach the same asymptotic value eventually.

Besides the two temporal scales ($T_t=2 \pi/\omega_t$, and $T_a=2 \pi/\omega_a$, see Figures \ref{imag_real} and \ref{imag_real_wake} in the Appendix B, where more details on the frequency jumps are given) associated to the frequency jumps, we can also observe a further periodicity, $T_f$ , related to the temporal modulation of the frequency during the early and intermediate terms (see Fig. 3c-d). This period is shorter ($T_f \sim 1$) for medium-short waves ($k > 10$) and longer ($T_f \sim 10^1-10^2$) for long waves ($k < 2$), and is in general different from $T_t$ and $T_a$. Moreover the system presents two other temporal scales: the external scale related to the base flow (see caption of Figures \ref{Fig_2} and \ref{Fig_3}) and the length of the transient (which can be determined by observing the time instant beyond which the growth rate, $r$, and the angular frequency, $\omega$, are both constant). Therefore, for each wavenumber, it is possible to count up to five different time scales.

Discontinuities on the frequency are well observable when transient dynamics are sufficiently extended in time. In general we observe that, for fixed wavenumbers, the transient behaviour for the channel flow lasts longer than for the wake flow. For both flows, short wavelengths lead to short transients, while long waves slowly extinguish their transient (for $k=1$ transients can last up to $10^3-10^4$ base flow time units, for $k=100$ only up to $10^{1}$ units). Moreover, for long waves in the wake, antisymmetric perturbations can in general present transients lasting longer than those observed for symmetric perturbations \cite{STC09}. However, as shown in Fig. \ref{sym_wake_frequency}, a different shape of symmetric initial conditions can lengthen transient dynamics. In synthesis, jumps of frequency are always clearly seen both for symmetric and antisymmetric perturbations in the channel and wake flows for all the longitudinal and oblique waves. The orthogonal waves that do not oscillate in time, and thus do not spatially propagate,  and that generally show the longest transient for a fixed wavenumber, have a frequency equal to zero at any time and thus  cannot  manifest  frequency jumps.

Temporal growth rates, $r$, are reported in panels (e) and (f) of Fig. 3 for the channel and wake flows, respectively. When transients monotonically grow or damp they are quite short and the temporal growth rates, $r$, become constant after few temporal scales (see the channel flow configurations in panel (e) and the symmetric perturbations for the wake flow in panel (f)). On the contrary, some transients can last thousands of time units. Examples of this are reported by the antisymmetric longitudinal perturbations acting on the plane wake flow (orange curves in panel f, $\phi = 0, \pi/4$). The temporal growth rates change their trend at about $t = 1000$ and $t = 1600$ for $\phi = 0$ and $\phi = \pi/4$, respectively. They still consistently increase beyond these points until the asymptotic states are reached ($t\simeq 3500-4000$ not reported in the Figure). The sudden variations of the temporal growth rates, $r$, are in correspondence with the instants where the frequency values, $\omega$, become constant and the amplification factors, $G$, change their trends (see panels d and b at about $t = 1000$ and t = $1600$).

A further comment can be made. Jumps in the frequency induce temporal acceleration or deceleration in the propagation speed of each wave.  This could  have an influence on the smoothness of the mean phase speed of a group of waves, in particular of the spots. For instance, in case the field contain multiple spots in various phases of their lives,  frequency jumps could  promote their  interaction because  acceleration-retardation of nearby ones gets them closer. As an example of multiple spot formation, see Figure 10 below, where the spot Temporal Building Window is 40 time scales long. In our opinion, in this concern, the concept of group speed as applied to the group of waves contained into a spot should be updated. Given the high time dependence of the frequency, the group velocity in this case becomes a time dependent variable. An update of this concept might be useful to interpret the complex propagation of the spots and  their forward and backward fronts.

\section{Discussion on transient dynamics}

\subsection{Intermediate transient}

\begin{figure}
\begin{center}
\includegraphics[width=0.7\columnwidth]{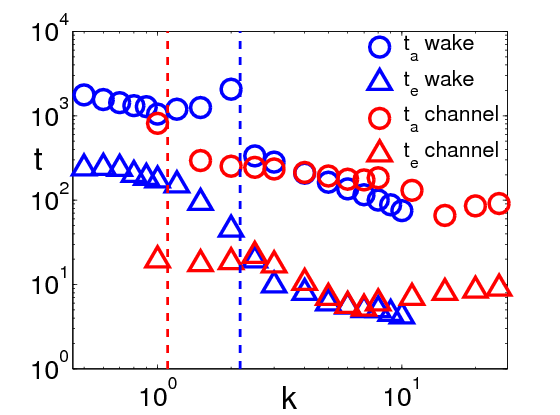}
\end{center}
	\caption{Typical transient time scales. $t_{a}$ (circles): time where the asymptotic limit is
reached ($r$ and $\omega$ settle to the final constant values). $t_{e}$ (triangles): time where the early transient ends and the frequency discontinuities occur. The intermediate term is given by the difference $t_a - t_e$, and generally is one order of magnitude longer than the early term. Blue symbols: wake flow, $Re=100$, antisymmetric initial conditions, $\phi=\pi/4$, the wake profile is observed at a distance from the body equal to 10 body scales, $x_0=10$, in a point located one body scale from the wake axis, $y_0=1$. Red symbols: channel flow $Re=10000$, symmetric initial conditions, $\phi=\pi/4$, the channel is longitudinally homogeneous, thus to specify the observation point it is sufficient to choose the transversal location, in this case the point is the midpoint between the wall and the channel axis, $y_0=0.5$. The dashed red and blue lines represent the transition from asymptotically unstable to stable wavenumbers.}
	\label{t_k}
\end{figure}

The observations of frequency jumps yield an interesting result: the perturbation temporal evolution has a three-part structure, with an early stage, an intermediate stage and an asymptotic stage. This is clearly seen by the fact that events like frequency jumps and associated fluctuations split the transient into two parts, where the second part is much longer than the first. We interpret these events as the beginning of the process that leads to the settlement of the asymptotic perturbation characteristics, that is the characteristics also predicted by the modal theory. The intermediate stage is the stage where this process takes place, while the early transient is the stage where the perturbation is most affected by the influence of the initial conditions. This observation should be framed in the general context of the 'intermediate asymptotics' where dynamical systems present solutions valid for times and distances from boundaries, large enough for the influence of the fine details of the initial /or boundary conditions to disappear, but small enough to keep the system far from the equilibrium state \cite{B66}.

\begin{figure}
\centering
\begin{minipage}[]{0.49\columnwidth}
\includegraphics[width=\columnwidth]{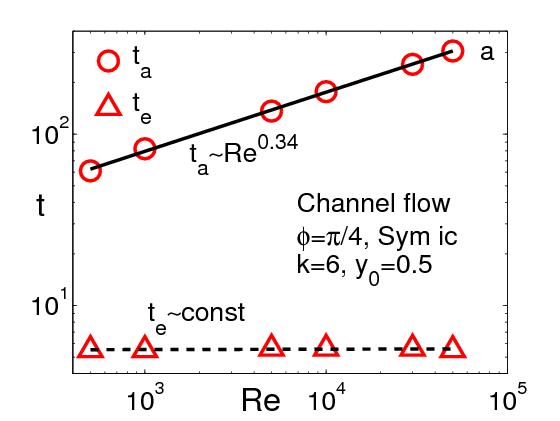}
\end{minipage}
\begin{minipage}[]{0.49\columnwidth}
\includegraphics[width=\columnwidth]{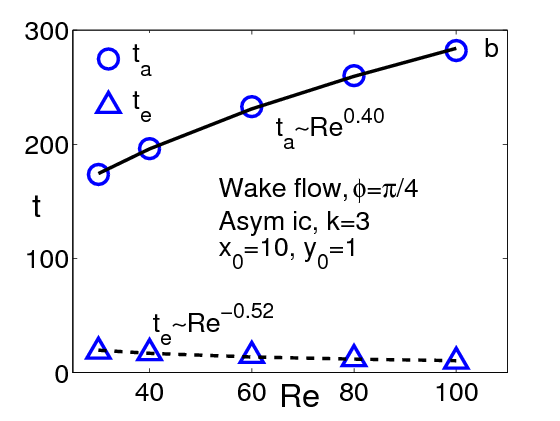}
\end{minipage}
\caption{Scaling of the transient time scales with respect to the Reynolds number: $t_{e}$ (triangles) is the time where the early transient ends and the frequency discontinuities occur, while $t_{a}$ (circles) is the time where the asymptotic limit is reached.  Panel (a): channel flow, $\phi=\pi/4$, symmetric initial conditions, $k=6$, the transversal observation point is $y_0=0.5$. $Re=[500, 50000$]. For the end of the mid-term we found a scaling $t_a \sim Re^{0.34}$ (solid curve), while the early transient remains constant, thus is not sensitive to the Reynolds number variation.  A log-log scale is adopted. Panel (b): wake flow, $\phi=\pi/4$, antisymmetric initial conditions, $k=3$, the longitudinal and transversal observation points are $x_0=10$ and $y_0=1$, respectively. $Re=[30, 100]$. The end of the early transient scales as $t_e \sim Re^{-0.52}$ (dashed curve), while the time where the asymptotic state is reached scales as $t_a \sim Re^{0.40}$ (solid curve). For both flows the intermediate region ($t_a - t_e$) increases with the Reynolds number. As an indication, the maximum relative error is, in the worst case, about $8\%$.}
\label{frequency_Re}
\end{figure}

The end of the early transient and the subsequent beginning of the intermediate transient is
announced by the occurrence of the frequency jumps. Many temporal scales beyond this instant, the frequency temporal variations disappear and a constant value emerges. The system, however, is not yet close to its ultimate state. The intermediate transient can
be considered extinguished only when the temporal growth rate, r, also becomes constant. At the transition between the early and intermediate terms, the perturbation suddenly changes its behaviour by varying its phase velocity. A measure of the temporal scales related to the end of the early transient and the reaching of the asymptotic state ($t_e$ and $t_a$, respectively) is reported in Fig. \ref{t_k}, by considering different perturbation wavelengths for both the wake and channel flows. The length of the intermediate transient can be obtained by calculating the difference between $t_a$ and $t_e$, and is in general one order of magnitude larger than the early term.

For two cases we are considering in this work, we have determined the scaling with the Reynolds number of the time where  the early part of the transient ends ($t_e$) and  the time where the transient ends and the  evolution becomes exponential ($t_a$), see the two figures below, \ref{frequency_Re}(a)-(b).
One  may notice that for the total transient duration, $t_a$,  the scaling presents positive exponents less than 1. The exponents for these oblique waves ($\phi = \pi/4$) are  close in the two cases (0.34 in the channel flow, 0.4 in the wake). The situation is different for the early transient time scale. The channel flow does not feel the Reynolds number variation, the wake instead presents a decay with exponent -0.52.  In any case, both cases evidence a definite trend of  growth for the intermediate term (equal to the difference $t_a - t_e$) with the Reynolds number. In general the intermediate term is more than one order of magnitude larger than the early transient.

In synthesis, our claim on the  tripartite structure of the temporal evolution of  travelling waves  is based on the  observation  of frequency jumps inside  the transients. The jumps split in two  the  life of the waves antecedent  the attainment  of the  exponential asymptote: (i) the initial stage (the early transient), heavily dependent on the initial condition, and (ii) a much longer stage, the intermediate transient,  which appears as a kind of intermediate asymptotics.  Since  scaling properties come on the stage when the influence of fine details of the  initial condition disappears but the system is still far from ultimate equilibrium state (intermediate asymptotics definition, see \cite{B66}), we propose the working hypothesis that the intermediate term we observe, either in case of unstable or stable waves,  should be nearly self-similar. In particular, we suppose to be in presence of a self-similarity of the second kind because we don't  think  that dimensional analysis in this case is sufficient for establishing self-similarity and scaling variables. This issue needs to be carefully considered and analyzed in future  dedicated studies.

In comparison with other problems of condensed matter, wave dynamics in dissipative systems is a relatively clean system whose lessons can be of greater value. Given the emphasis on similar topics in geophysics,  pattern formation, MHD and plasma dynamics,  we think that the question of whether a universal state (in this case the intermediate term)  exists independently of the forcing  is a  typical  issue for the research in several of these areas.

\subsection{Orthogonal waves}

Disturbances normal to the mean flow ($\phi=\pi/2$) do not oscillate in time, thus have zero frequency and phase velocity throughout their lives.  This means  that orthogonal waves do not propagate. Indeed, by symmetry there is no reason for an orthogonal wave to move in either of the two possible directions along the $z$ coordinate. In fact,  the base flows here considered do not have a component in this  direction. On the contrary, the phase velocity is maximum for longitudinal waves because these have the same direction of the base flow.

\begin{figure}
\begin{center}
\includegraphics[width=0.7\columnwidth]{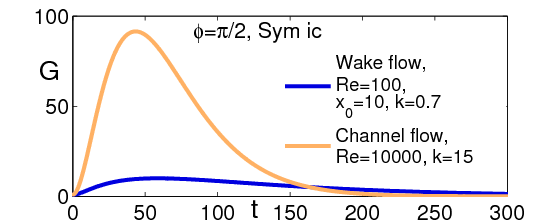}
\end{center}
	\caption{Examples of transient growths for waves orthogonal ($\phi=\pi/2$) to the mean flow, symmetric initial conditions. Wake flow: $Re=100$, $x_0=10$ diameters downstream the body, $k=0.7$. Channel flow: $Re=10000$, $k=15$. The channel transient growths are in general much more intense than those observed in the wake flow. For a visualization of the growth-decay of the orthogonal standing wave in the channel flow, see the film \textsf{ChannelStandingWave.avi}.}
	\label{orthogonal_transient}
\end{figure}

Orthogonal waves, although always asymptotically stable, can experience a quick initial growth of energy for high enough Reynolds numbers. This behaviour is evidenced in Fig. \ref{orthogonal_transient}, where two examples of transient growths are reported for the channel and wake flows.  Both configurations are asymptotically stable but, before this state is reached, these waves have strong amplifications, which can last up to hundreds of time units. It should be recalled that the maximum growth increases reducing the wavenumber, for instance for the channel flow (Re=10000) when $k=2$ amplification factor values as high as $10^4$ can be observed. On the basis of these findings, the role of orthogonal waves, often underestimated due to their asymptotic stability, can be considered important for the understanding of mechanisms such as  nonlinear wave interaction and bypass transition \cite{L1996,HLJ1993}. It should be also recalled that the laboratory images that describe turbulent spots \cite{CCD1978,GBR1981,H1996} show in their back part longitudinal streaks  which are not travelling  across the channel. This pattern can be associated to the excitation of an orthogonal wave and to its particularly long transient \cite{STC09,CJLJ97}.  In fact, spots are transitional structures that last for a long time inside the flow, which has made them observable in the laboratory since many decades \cite{GBR1981,KTS1981,CCD1978}. The duration of spots is not well documented in literature, however what is known does not seem incompatible with the typical length of orthogonal wave long transients.  All this can thus offer a possible interpretation for the formation and the morphology of transitional spots.

\subsection{Spot formation by linear superposition of orthogonal and travelling waves}

In both the channel and the wake flows, we tried  to reconstruct the formation of a wave packet   centered around a  given  wave-number value. This is carried out  by superposing around an observation  point quite a large number  of three dimensional perturbation waves. The waves are not exactly in phase because they enter stochastically  the system in a temporal building window (TBW)  with an extension of many temporal scales.  The assumption here is that the sum in a spatial domain of a large set of  plane perturbations  in a narrow range of wave-numbers and stochastically out of phase  is an approximation of the general three dimensional  linear perturbation localized  about some reference point.
For both flows, we have selected a certain wave number, $k$, with a  surrounding, $dk$, and computed the perturbation lives of about 360 waves distributed  over 36 obliquity angles in the range $[-\pi/2, \pi/2]$, 5 $k$ values in the $[k-dk, k+dk]$ range and 2 initial arbitrary conditions (symmetric and antisymmetric). We have then randomly superimposed half of them in the observation point over a  time interval, the building window, equal to 10 or 40 physical time scales. The instant where a randomly selected  wave enters the sum is stochastically chosen inside the  building interval. The probability distribution of the choices is uniform.

The temporal evolution of the packet can be followed up to the final decay which ends when the longest transient in the packet will die out.
The only non randomly selected wave is the orthogonal wave with wave-number $k$ which is introduced at the initial instant.
The triggering of the packet is imagined as associated to the excitation  at a certain  instant of an orthogonal wave. This because: (i)  in all the laboratory and numerical images representing spots in the Couette, boundary layer and plane Poiseuille flows the back part of the spot always contains evidence of wave crests and valley parallel to the basic stream-wise direction (see, among many others, \cite{CCD1978,GBR1981,H1996}; (ii) the orthogonal waves, even if asymptotically  always stable, very often presents very large transient growth. And when this is does not happen, as at very low subcritical Reynolds numbers, they in any case  are the least stable waves.

Furthermore, the overall transient length is maximum for the orthogonal waves, which differently from all the other waves,  are not travelling waves. In fact, it is important to recall that they are standing waves (the angular frequency $\omega$ and the phase speed are identically zero) though transiently growing or decaying. See the film \textsf{ChannelStandingWave.avi} in the Supplementary Material uploaded in the NJP online repository.

The  longitudinal wave  has the largest phase speed while the oblique waves show a cosine variation of the phase speed with the obliquity angle (see figure  \ref{frequency_orthogonal}). Thus, when considering the path covered by the fan of perturbation crests in a given time interval, it can be deduced that the  visible borders of the spot pattern must be roughly heart shaped (see, for instance, \cite{CCD1978,CWP1982,DHB1992}).

The formation and evolution of two spots in the Poiseuille flow, and one in the wake, can be seen in figures 9, 10, 11.

The fact that the subcritical spot in the wake forms in a very residual way, when all the waves in the packet are close to die out, may explain why they were not yet seen in the laboratory. Our visualization can be compared with many images in literature, see e.g. \cite{DHB1992} Fig. 2(d - e - f), \cite{HLJ1993} Fig. 2b,
\cite{DSH2010} Fig. 4a, \cite{CWP1982} Fig. 4, \cite{DD1995} Fig. 6, \cite{CCD1978} Fig. 6a, \cite{H1996} Fig. 3a, \cite{TTA2010} Fig. 6a. All the spot images cited  are laboratory or  Navier-Stokes DNS, where the nonlinear effects are included, but the linear DNS image by \cite{HLJ1993}. One  may notice a good qualitative comparison between our results and the others in literature. It should be considered that our spots (see also in the online Supplementary Material  the films \textsf{channel\_50fps.avi} and \textsf{wake\_50fps.avi}) are young and may  represent only the beginning  of the transient  lives. It should be recalled that to obtain an efficient determination of the perturbation solution  an adaptive technique must be implemented. To produce  the film, all the 186 waves used to build the spots must be summed up from the instant each of them  enters the system. To compute the sum all the waves  must be interpolated  on the same  regular time steps.  The film production is thus very long and cumbersome. We have produced at the moment 40 time scales. On the contrary, visualizations in literature usually show rather old spot, i.e. pattern  hundred of time scales old. Furthermore,  in our visualization, the nonlinear  coupling  is missing. This  circumvents the formation of the wiggles due to the coupling of the oblique and longitudinal waves with the orthogonal one.  These wiggles are always observed in the spot forward  parts  in the laboratory and  nonlinear simulation visualizations.

\begin{figure}
\centering
\begin{minipage}[]{0.49\columnwidth}
\includegraphics[width=\columnwidth]{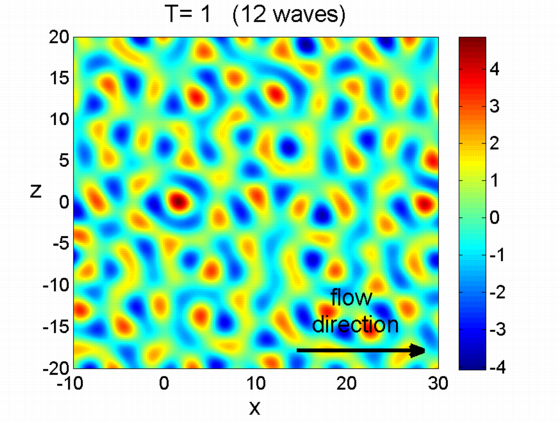}
\end{minipage}
\begin{minipage}[]{0.49\columnwidth}
\includegraphics[width=\columnwidth]{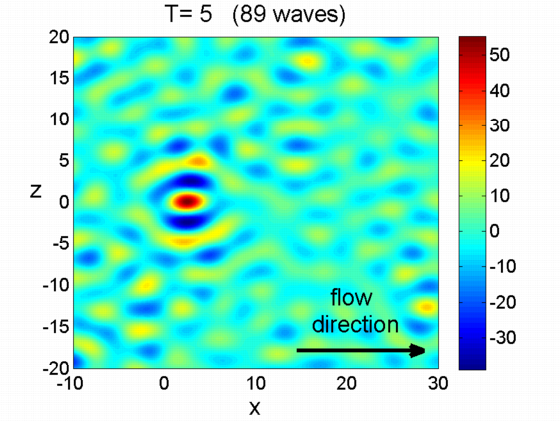}
\end{minipage}
\begin{minipage}[]{0.49\columnwidth}
\includegraphics[width=\columnwidth]{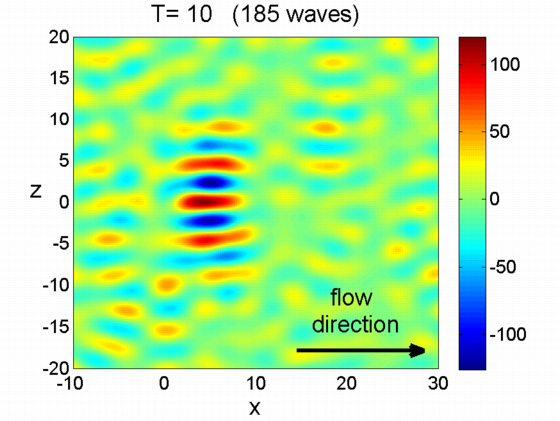}
\end{minipage}
\begin{minipage}[]{0.49\columnwidth}
\includegraphics[width=\columnwidth]{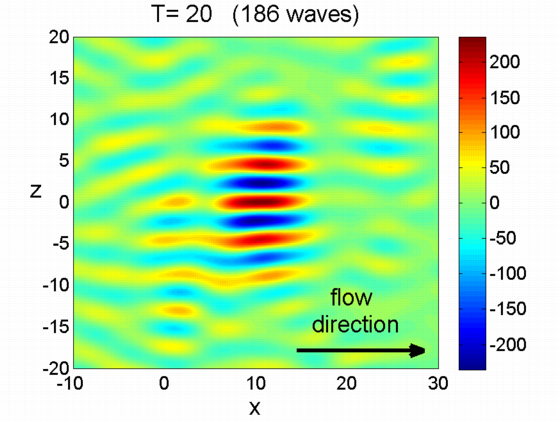}
\end{minipage}
\begin{minipage}[]{0.49\columnwidth}
\includegraphics[width=\columnwidth]{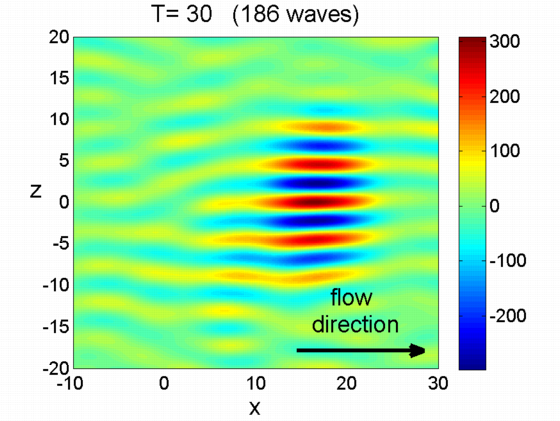}
\end{minipage}
\begin{minipage}[]{0.49\columnwidth}
\includegraphics[width=\columnwidth]{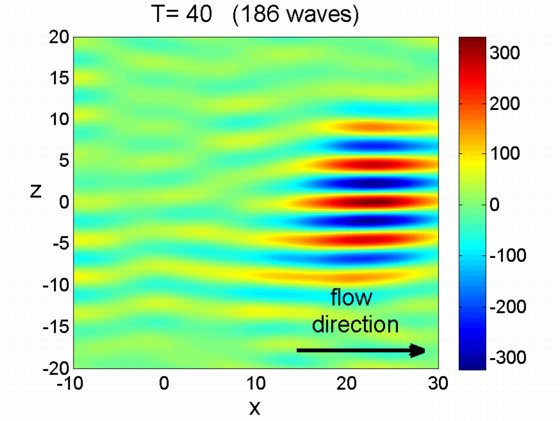}
\end{minipage}
\caption{Poiseuille flow, $Re=1500$. Spot formation. Stochastic superposition of 186 waves in a  temporal building window lasting 10 times scales, see text above. The Reynolds number is 1500, the wave-number central value  is $1.4 \pm 0.2$ . The stream-wise  velocity, u,  is visualized in the physical plane (x,z) at different temporal instants and at  the transversal  point, $y_0 = 0.5$. The visualization is made taking about 157 points over the shortest wavelength (k=1.6) and 239 temporal intervals over the shortest temporal period $(\omega = 1.05)$. Please, see also the associated film \textsf{channel\_50fps.avi}, included in the online Supplementary Material. By observing the film, we have measured an average velocity for the spot center nearly equal to 0.6 speed units. This is very close to the Carlson et al. result at Re=1000 \cite{CWP1982}: the front of the spot moves with a propagation velocity of about $O.6U$, while the rear moves at a propagation velocity of $0.34U$ (U is the mean speed in the channel center).
}
\end{figure}

\begin{figure}
\centering
\begin{minipage}[]{0.49\columnwidth}
\includegraphics[width=\columnwidth]{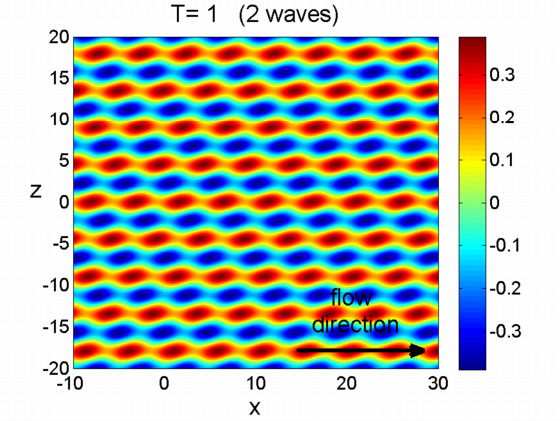}
\end{minipage}
\begin{minipage}[]{0.49\columnwidth}
\includegraphics[width=\columnwidth]{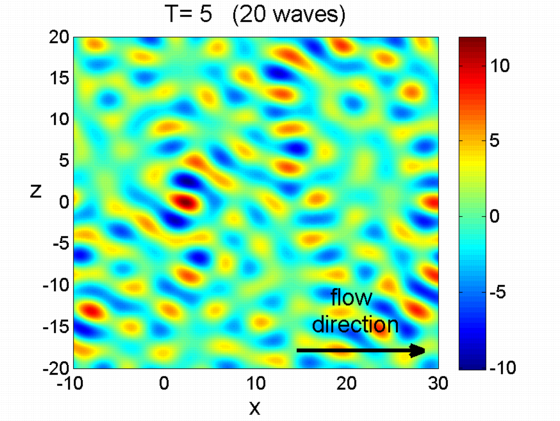}
\end{minipage}
\begin{minipage}[]{0.49\columnwidth}
\includegraphics[width=\columnwidth]{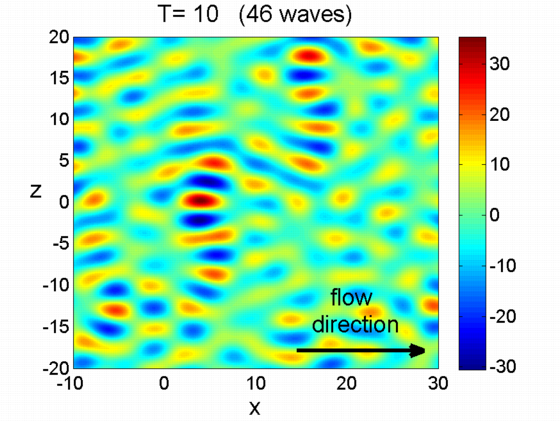}
\end{minipage}
\begin{minipage}[]{0.49\columnwidth}
\includegraphics[width=\columnwidth]{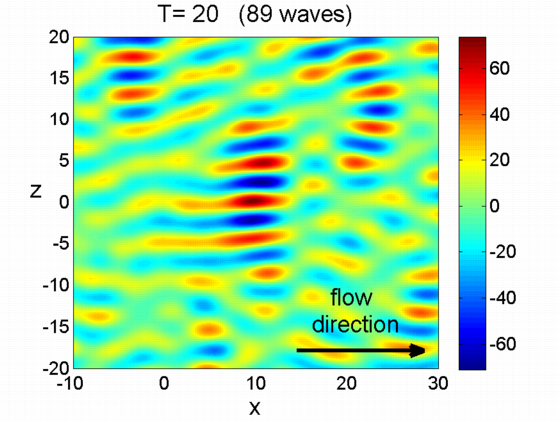}
\end{minipage}
\begin{minipage}[]{0.49\columnwidth}
\includegraphics[width=\columnwidth]{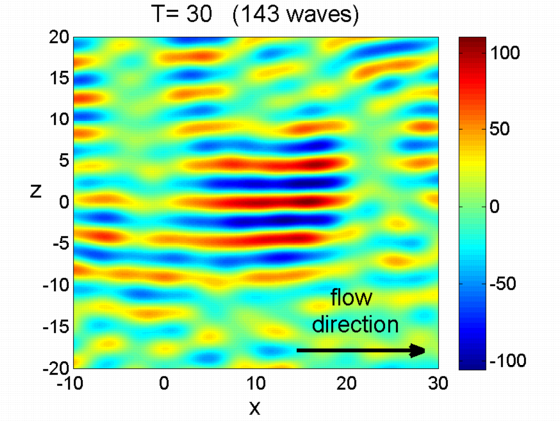}
\end{minipage}
\begin{minipage}[]{0.49\columnwidth}
\includegraphics[width=\columnwidth]{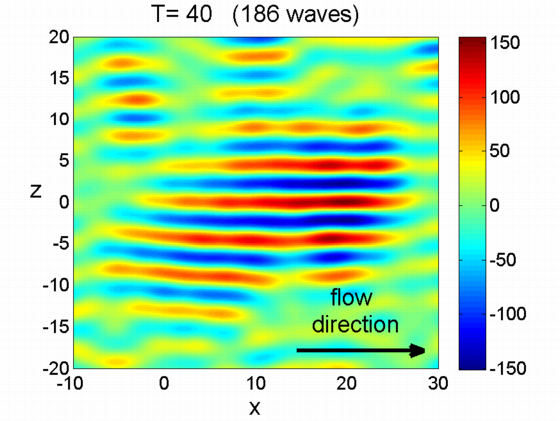}
\end{minipage}
\caption{
Spot formation in the channel flow. Everything is as described in the caption of the previous figure but the temporal building window which  here lasts 40 times scales. The superposed waves are the one with respect to the other statistically less in phase.  This produces a more complex pattern where the onset of  multiple spots can be observed}
\end{figure}

\begin{figure}
\centering
\begin{minipage}[]{0.49\columnwidth}
\includegraphics[width=\columnwidth]{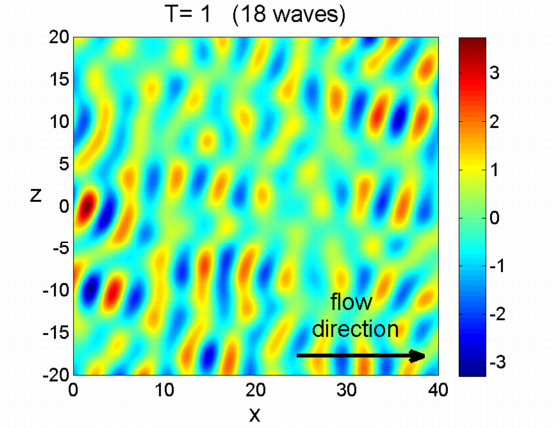}
\end{minipage}
\begin{minipage}[]{0.49\columnwidth}
\includegraphics[width=\columnwidth]{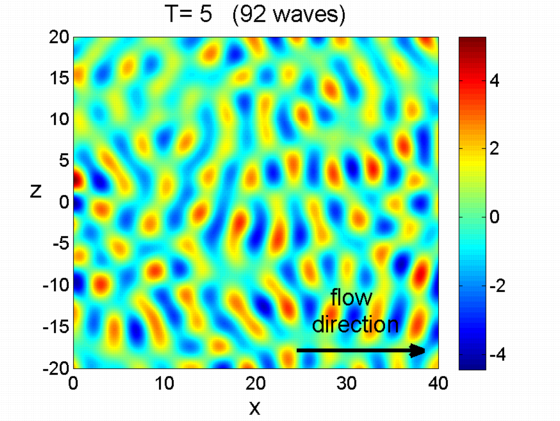}
\end{minipage}
\begin{minipage}[]{0.49\columnwidth}
\includegraphics[width=\columnwidth]{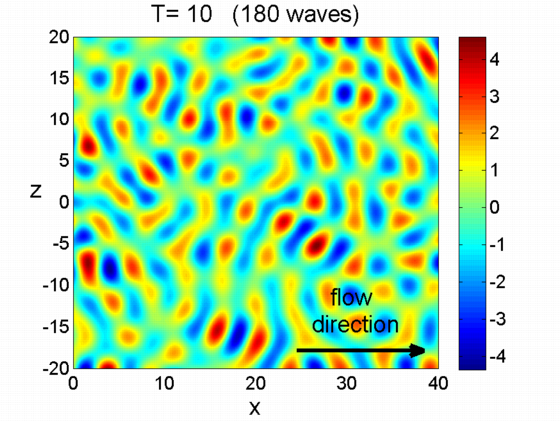}
\end{minipage}
\begin{minipage}[]{0.49\columnwidth}
\includegraphics[width=\columnwidth]{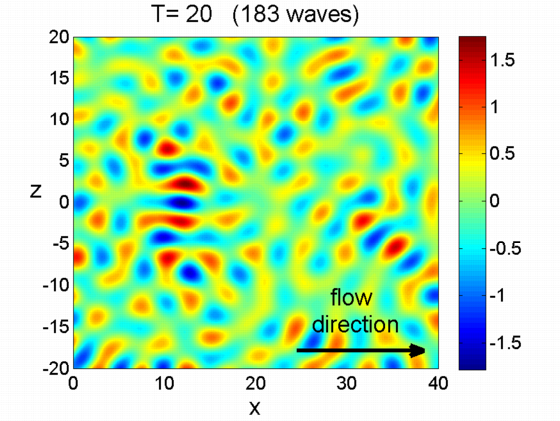}
\end{minipage}
\begin{minipage}[]{0.49\columnwidth}
\includegraphics[width=\columnwidth]{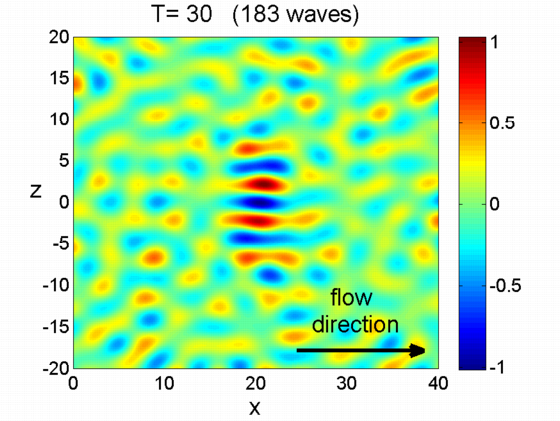}
\end{minipage}
\begin{minipage}[]{0.49\columnwidth}
\includegraphics[width=\columnwidth]{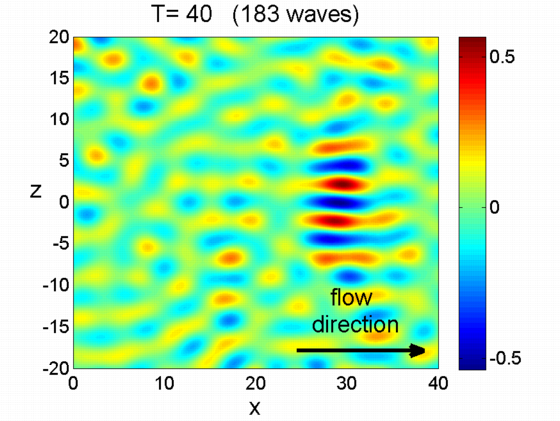}
\end{minipage}
\caption{Plane wake flow, $Re=30$, $x_0=50$.
Spot formation. Stochastic superposition of 183 waves in a  temporal building window lasting 10 times scales, see text above. The Reynolds number is 30, which is subcritical, the wave-number central value  is $1.5 \pm 0.2$. The section of the wake here considered is placed  50 scale units downstream the body. The stream-wise velocity, $u$, is visualized in the physical plane $(x,z)$ at different temporal instants and at a  transversal position $y_0$ equal to 1. The visualization is made taking about 148 points over the shortest wavelength $(k=1.7)$ and 173 temporal intervals over the shortest temporal period ($\omega = 1.45$). Here, the spot forms later than in the channel case. The orthogonal wave does not have a transient growth, but is the least stable. It takes a certain time lag to overcame in amplitude the other waves. When this happens the spot is visible. However, it is very residual in intensity. Please, confer the intensity of the channel spots and consider that the legend aside each panel in these visualizations is renormalized on the instantaneous  maximum value of the stream-wise velocity. Please, see the film \textsf{wake\_50fps.avi}, which is  included in the online Supplementary Material.
}
\end{figure}

\section{Some information on the asymptotic behaviour of the dispersion relation}

\begin{figure}
\centering
\begin{minipage}[]{0.49\columnwidth}
\includegraphics[width=\columnwidth]{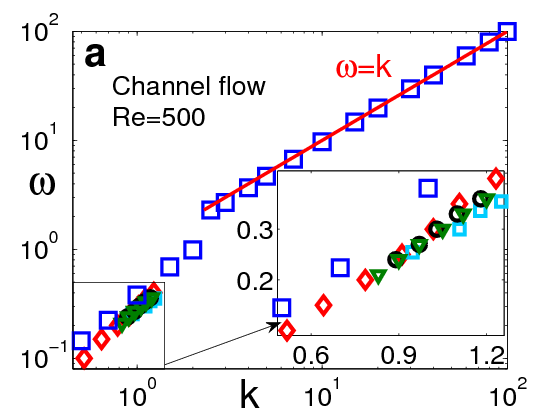}
\end{minipage}
\begin{minipage}[]{0.49\columnwidth}
\includegraphics[width=\columnwidth]{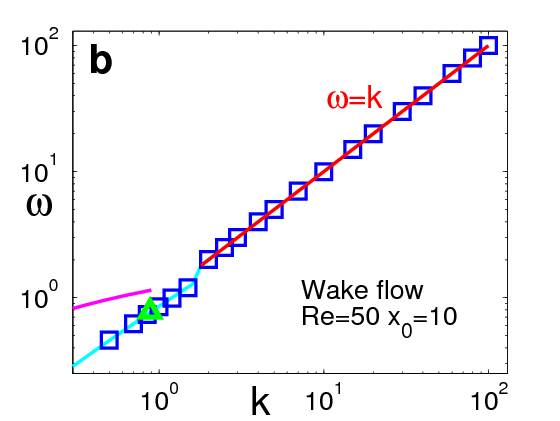}
\end{minipage}
\begin{minipage}[]{0.49\columnwidth}
\includegraphics[width=\columnwidth]{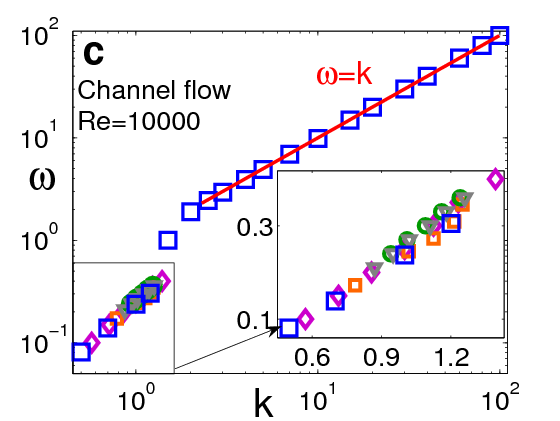}
\end{minipage}
\begin{minipage}[]{0.49\columnwidth}
\includegraphics[width=\columnwidth]{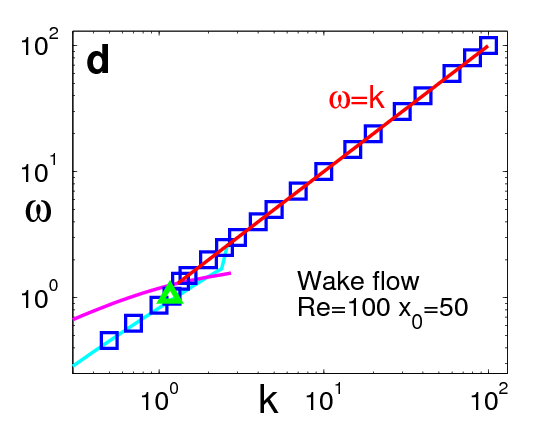}
\end{minipage}
\begin{center}
\vspace{-0.2cm}
\includegraphics[width=0.8\columnwidth]{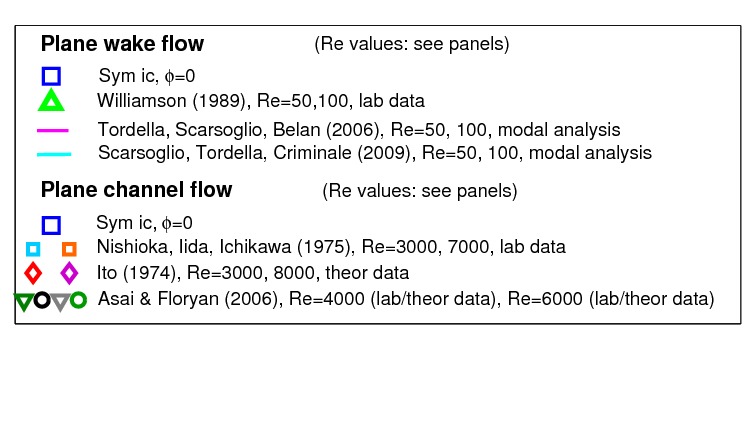}
\end{center}
\caption{Spectrum in the wavenumber space of the asymptotic frequency for a collection of longitudinal waves ($\phi = 0$). The plane channel flow is on the left ($Re = 500$ top panel, $Re = 10000$ bottom panel), the bluff body plane wake is on the right ($Re = 50$ and $x_0 = 10$ top panel, $Re = 100$ and $x_0 = 50$ bottom panel). The observation transversal points are $y_0 = 1$ and $y_0 = 0.5$ for the wake and channel flows, respectively. Symmetric initial conditions are indicated by blue squares. The data in this figure are compared to the available laboratory and theoretical data where frequencies are associated to specific wavenumber values, see legend (references \cite{Ni1975,Ito1974,Asai2006,W89,TSB2006,ETC12}).}
\label{frequency}
\end{figure}

In this section we describe the distribution of the frequency and phase velocity of longitudinal and transversal waves in correspondence to the settlement of the asymptotic condition. As mentioned in Section 2, this condition can be considered reached when both the temporal growth rate, $r$, and the angular frequency, $\omega$, approach a constant value. It should be noted that in all the cases we observed, the frequency settles before the growth rate.

\subsection{Frequency spectral distribution of longitudinal waves}

The frequency determination can be validated through the comparison of the temporal asymptotic behaviour obtained by means of the initial-value analysis with other theoretical and experimental data in literature. To our knowledge, this data collection, unfortunately, does not contain information on three-dimensional perturbations. Indeed, the normal mode theory is quite restricted to longitudinal perturbations. For the channel we refer to the available results in \cite{Ni1975,Ito1974,Asai2006}, for the wake, to the results in \cite{W89,TSB2006,ETC12}. The asymptotic frequency dependence on the wavenumber is presented in spectral form in Fig. \ref{frequency}. We observe a good agreement  between the different literature data and the present results. It should be noted that this is much so for long waves, the most unstable ones \cite{STC10}. Indeed, these perturbations are those easily observed in the laboratory, even if, usually in their nonlinear regime. We see that although experimental results are affected by the nonlinear interaction, the agreement between laboratory data and linear IVP analysis is very good. It has been shown \cite{DelCho1998} that nonlinear terms limit the amplitude of the wave packet, leaving unaffected its frequency, see also the laboratory and normal mode data comparison in \cite{TSB2006,BT2006}. This good data agreement validates the use of linear stability analysis to predict the frequency transient and asymptotic behaviour.

Figure \ref{frequency} presents an extended spectral dependence of the frequency, $\omega$, on the polar wavenumber, $k$ (a log-log scale is adopted). In fact, it contains more than two decades of wavenumbers ($k \in [0.5, 100]$), which is uncommon in literature. For longitudinal waves ($k=\alpha$), see Fig. 1 a, and large enough wavenumber values (about $k>2$) we observe that $\omega=k$ (see red curves in Fig. \ref{frequency}), for both the base flows and the configurations here considered ($Re=500, 10000$ for the plane channel flow, $Re=50, 100$ for the plane wake flow). This means that the behaviour is non-dispersive. For smaller wavenumbers, instead, $c$ is a complicated function of $k$ and the behaviour becomes dispersive. The transition between dispersive and non-dispersive behaviour is highlighted in Fig. \ref{frequency}, where sudden variations of the spectral frequency distribution, $\omega(k)$, occur in the surroundings of $k=2$. Jumps in the asymptotic spectral distribution are more marked for the channel flow (panels (a) and (c)) rather than for the wake flow (panels (b) and (d)). The transition between dispersive and non-dispersive regions is even more evident if one considers the spectral distribution of the amplitude of the phase velocity, $|C(k)|$, as shown in Fig. \ref{frequency_orthogonal} of the next Section.

\subsection{Frequency and phase speed for oblique perturbations}

\begin{figure}
\centering
\begin{minipage}[]{0.49\columnwidth}
\includegraphics[width=\columnwidth]{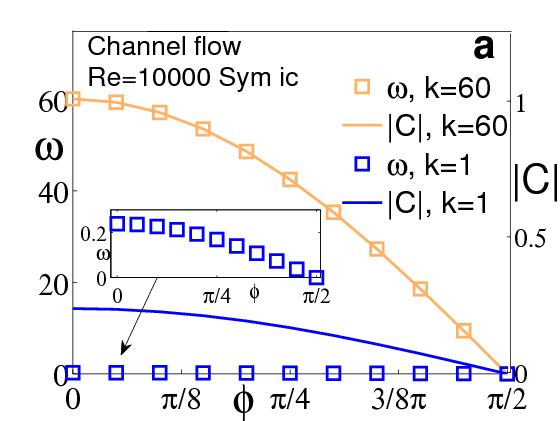}
\end{minipage}
\begin{minipage}[]{0.49\columnwidth}
\includegraphics[width=\columnwidth]{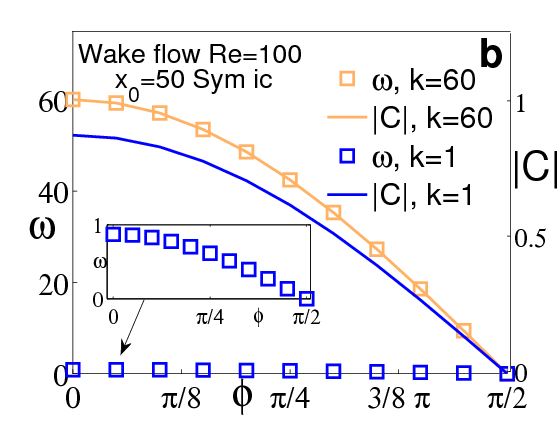}
\end{minipage}
\begin{minipage}[]{0.49\columnwidth}
\includegraphics[width=\columnwidth]{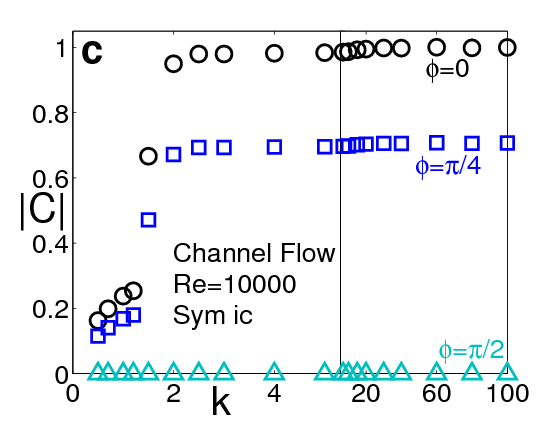}
\end{minipage}
\begin{minipage}[]{0.49\columnwidth}
\includegraphics[width=\columnwidth]{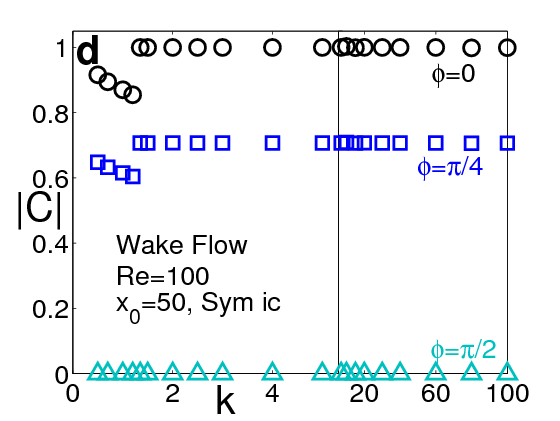}
\end{minipage}
\caption{Asymptotic frequency, $\omega$, and phase speed module, ${\bf |C|}$, as functions of the angle of obliquity, $\phi$, and of the perturbation wavelength. (a) channel flow $Re=10000$, symmetric initial conditions, $k=1, 60$, (b) wake flow, $Re=100$, symmetric initial conditions, $k=1, 60$. In panels (c)-(d) the asymptotic spectral distribution of the phase velocity module is presented for three angles of obliquity ($\phi=0, \pi/4, \pi/2$) and $0.5<k<100$. (c) channel flow, $Re=10000$, symmetric initial conditions; (d) wake flow, $Re=100$, symmetric initial conditions. The observation points are $y_0 = 1$ and $x_0=50$ for the wake, and $y_0 = 0.5$ for channel flow.}
\label{frequency_orthogonal}
\end{figure}

In Fig. \ref{frequency_orthogonal}(a)-(b) the frequency, $\omega$, and the module of the phase velocity, $|\textbf{C}|$, are reported as functions of the obliquity angle, $\phi$, for two different polar wavenumbers, $k$. For both  the flows, the ordinate axis on the left represents the frequency, $\omega$, while the one on the right, the module of the phase velocity, ${\bf |C|}$. We verified that for short waves

\begin{equation}
\omega=k\ c\ \cos(\phi),
\end{equation}

\noindent where $|\textbf{C}| = c\ \cos(\phi)$ is the module of the phase velocity. For longer waves ($k<1.5$ for the wake, $k<2$ for the channel flow), we observe that $c(k)$ is  highly dependent on $k$ and on the basic flow. Thus,

\begin{equation}
\omega=k\ c(k)\ \cos(\phi),
\end{equation}

\noindent and the phase velocity vector, ${\bf C}$, has components

\begin{equation}
C_x = c(k) \cos^2(\phi), \,\,\,\, C_z = c(k) \cos(\phi) \sin(\phi),
\end{equation}

In Fig. \ref{frequency_orthogonal}(c)-(d), the module, ${\bf |C|}$, is shown both for the channel and wake flows as a function of $k$ for three different angles of obliquity. Non-orthogonal long waves are dispersive as the shape of $c(k)$ strongly depends on $k$. It should be also noticed that the dispersion higly depends on the base flow considered. In fact, the phase speed variation in the wall flow is opposite to that in the free flow. For shorter waves, $|C|$ approaches a constant value which only depends on the angle of obliquity, $\phi$.

\section{Conclusions}

This study presents a newly observed phenomenology  relevant to  shear flows perturbation waves.
In two different archetypical  shear flows,  the  plane  channel  and the plane wake flows, and for two Reynolds numbers  (500 and 10000 in the channel,  50 and 100 in the wake), we yield empirical evidence that transient solutions of travelling waves at any wave number inside the range $k \in [0.5, 100]$, and any obliquity angle have a tripartite structure. This is composed by an early, an intermediate and a long term. This last starts when  the asymptotic exponential behaviour is reached.  %Given this, we may consider it possible to suppose that  this phenomenon is a general feature of the  solutions  of the linear   Navier-Stokes equations.

The claim on the  tripartite structure of the temporal evolution of  travelling waves is based on the  observation  of frequency jumps inside  the transients. The jumps split in two  the  life of the waves antecedent  the attainment  of the  exponential asymptote: a first part, the early transient, which is  heavily dependent on the initial condition, and  a second much longer part, the intermediate transient,  which appears as a kind of intermediate asymptotics.  On the basis of the fact that scaling comes on a stage when the influence of fine details of the  initial condition disappears but the system is still far from ultimate equilibrium state, we advance the hypothesis  that the intermediate term we observe, either in case of unstable or stable waves,  should be nearly self-similar. In particular, we suppose to be in presence of a self-similarity of the second kind because we don't  think  that dimensional analysis in these systems is sufficient for establishing self-similarity and scaling variables. Of course, this issue needs to be carefully considered and analyzed in future  dedicated studies.

%To enter a bit in details, %We show evidence of a discontinuous behaviour in the frequency within the transient life of three-dimensional travelling perturbation waves in two typical sheared flows.
%In the wall flow case, the presence of the frequency discontinuity is barely influenced by the symmetry of the initial condition or by the obliquity angle of the perturbation wave. In the free flow, the plane wake, the jumps are more marked for antisymmetric perturbations, either aligned with the basic flow or oblique to it.
The jumps appear in the evolution after many basic flow eddy turn over times have elapsed. The duration of the early term is typically not larger than about the 10\% of the global transient length. As a consequence, since after the intermediate term perturbations die out or blowup, the mid-term can be considered as the most probable state in the life of a perturbation.

In general, frequency jumps are preceded by a modulation of the frequency value observed in the early transient and followed by higher or lower values with a modulation that progressively extinguishes as the asymptotic state is approached.  The only waves which do not show frequency discontinuities are the orthogonal waves. These do not propagate and, in subcritical situations, are asymptotically the least stable.

Jumps in the frequency induce temporal acceleration or deceleration in the propagation speed of a perturbation wave.  This could  have an influence on the smoothness of the mean phase speed of a group of waves, in particular of the waves in the spots commonly observed in transitional wall flows. For instance, in case the field contain multiple spots in various phases of their lives,  frequency jumps could  promote their  interaction because  acceleration-retardation of nearby ones can get them closer.

In both the channel and the wake flow, we tried  to simulate the formation of a wave packet centered around a given  wave-number value. This was carried out  by superposing about an observation  point quite a large number  of three dimensional perturbation waves. The waves are not in phase because they enter stochastically the system in a temporal building window which lasts many eddy turn over times. At the initial instant, the superposition starts with  an orthogonal wave. The assumption here is that the sum of a large set of three-dimensional perturbations,  in a narrow range of wave-numbers and sufficiently out of phase, is an acceptable approximation of a general localized three dimensional small perturbation.

In both flow cases, the linear initial stage of a typical spot formation characterized by longitudinal streaks was observed. In the subcritical channel flow (Re = 1500)
the spot intensity is growing quickly  due to the contribution coming from the transient growth of the orthogonal wave. Instead, in the subcritical wake (Re = 30),
the spot forms in a very residual way when all the waves in the packet are close to die out and the contribution from the orthogonal one (the least stable) is eventually prevailing. This can explain why  spots in the wake will hardly enter the nonlinear stage and are not  seen in the laboratory.

The investigation of the dispersion relation in the asymptotic regime reveals that long waves, both longitudinal and oblique, with a wavenumber below 2 in the channel flow and 1.5 in the wake flow, present a dispersive behaviour. If, inside a spot, the wavenumber is distributed in a narrow but finite range inside the dispersion region, in the long term the longest wave will propagate in a different way than the shortest one. This can explain the spot spatial growth.

\section*{Acknowledgments}
\addcontentsline{toc}{section}{Acknowledgments}

The authors thank Ka-Kit Tung, William O. Criminale, Miguel Onorato and Davide Proment for fruitful discussions on the results presented in this work.

%\appendix

\section*{Appendix A. Initial-value problem formulation}
\addcontentsline{toc}{section}{Appendix A. Initial-value problem formulation}

The base flow system is perturbed with small three-dimensional disturbances. The perturbed system can be linearized and the continuity and Navier-Stokes equations describing its spatio-temporal evolution can be expressed as:

\begin{equation}
\frac{\partial \widetilde{u}}{\partial x} + \frac{\partial \widetilde{v}}{\partial y} + \frac{\partial \widetilde{w}}{\partial z} = 0, \\ \label{IVP2_continuity}
\end{equation}
\begin{equation}
\frac{\partial \widetilde{u}}{\partial t} + U \frac{\partial \widetilde{u}}{\partial x} + \widetilde{v} \frac{dU}{dy} + \frac{\partial \widetilde{p}}{\partial x} = \frac{1}{Re} \nabla^2\widetilde{u}, \\ \label{IVP2_NS1}
\end{equation}
\begin{equation}
\frac{\partial \widetilde{v}}{\partial t} + U \frac{\partial \widetilde{v}}{\partial x} + \frac{\partial \widetilde{p}}{\partial y} = \frac{1}{Re} \nabla^2\widetilde{v},\\
\label{IVP2_NS2}
\end{equation}
\begin{equation}
\frac{\partial \widetilde{w}}{\partial t} + U \frac{\partial \widetilde{w}}{\partial x} + \frac{\partial \widetilde{p}}{\partial z} = \frac{1}{Re} \nabla^2\widetilde{w} \label{IVP2_NS3}
\end{equation}

\noindent where ($\widetilde{u}(x, y, z, t)$, $\widetilde{v}(x, y,
z, t)$, $\widetilde{w}(x, y, z, t)$) and $\widetilde{p}(x, y, z,
t)$ are the perturbation velocity and pressure, respectively. $U$ and $dU/dy$ indicate the base flow profile (under the near-parallelism assumption) and its first derivative in the shear direction, respectively. For the channel flow, the independent spatial variable, $z$, is defined from $-\infty$ to $+\infty$, the $x$ variable from $-\infty$ to $+\infty$, and the $y$ from $-1$ to $1$. For the plane wake flow, $z$ is defined from $-\infty$ to $+\infty$, $x$ from $0$ to $+\infty$, and $y$ from $-\infty$ to $+\infty$. All the physical quantities are normalized with respect to a typical velocity (the free stream velocity, $U_f$, and the centerline velocity, $U_0$, for the 2D plane wake and the plane Poiseuille flow, respectively), a characteristic length scale (the body diameter, $D$, and the channel half-width, $h$, for the 2D plane wake and the plane Poiseuille flow, respectively), and the reference density, $\rho_0$.

The plane channel flow is homogeneous in the $x$ direction and is represented by the Poiseuille solution, $U(y)=1-y^2$. Assuming that the bluff-body plane wake slowly evolves in the streamwise direction, the base flow is approximated at each longitudinal station past the body, $x_0$, by using the first orders ($n = 0, 1$) of the Navier-Stokes expansion solutions described in \cite{TB03}. Under this approximation, $U(y; x_0, Re)=1 - a x_0^{-1/2} \rm
\exp \left(- \frac {Re}{ 4} \frac {y^2}{ x_0}\right)$, where $a$ is related to the drag coefficient.

By combining equations (\ref{IVP2_continuity}) to (\ref{IVP2_NS3}) to eliminate the pressure terms, the perturbed system can be expressed in terms of velocity and vorticity \cite{CD90}. A two-dimensional Fourier transform is then performed in the $x$ and $z$ directions for perturbations in the channel flow. Two real wavenumbers, $\alpha$ and $\gamma$, are introduced along the $x$ and $z$ coordinates, respectively. A combined two-dimensional Laplace-Fourier decomposition is instead performed for the wake flow in the $x$ and $z$ directions. In this case, a complex wavenumber, $\alpha = \alpha_r + i \alpha_i$, is introduced along the $x$ coordinate, as well as a real wavenumber, $\gamma$, along the $z$ coordinate. To obtain a finite perturbation kinetic energy, the
imaginary part, $\alpha_i$, of the Laplace transformed complex longitudinal wavenumber
can only assume non-negative values and can thus be defined as a spatial damping rate in the streamwise direction.
Here, for the sake of simplicity, we have $\alpha_i=0$, therefore $\alpha=\alpha_r$. The following governing partial
differential equations are thus obtained

\begin{eqnarray} \label{IVP2_fou1}
\frac{\partial^2 \hat{v}}{\partial y^2} &-& k^2 \hat{v} = \hat{\Gamma}, \\
\frac{\partial \hat{\Gamma}}{\partial t} &=& - i k \cos(\phi) U \hat{\Gamma} + i k \cos(\phi) \frac{d^2 U}{dy^2} \hat{v} + \frac{1}{Re} \left(\frac{\partial^2 \hat{\Gamma}}{\partial y^2} -
k^2 \hat{\Gamma}\right),\label{IVP2_fou2} \\
 \frac{\partial \hat{\omega}_{y}}{\partial t} &=& - i k \cos(\phi) U \hat{\omega}_{y} - i k \sin(\phi) \frac{dU}{dy} \hat{v} + \frac{1}{Re} \left(\frac{\partial^2 \hat{\omega}_y}{\partial y^2} - k^2
\hat{\omega}_y\right),\label{IVP2_fou3}
\end{eqnarray}

\noindent where the superscript $^\wedge$ indicates the transformed perturbation quantities. The quantity $\hat{\Gamma}$ is defined through the kinematic relation $\widetilde{\Gamma} = \partial \widetilde{\omega}_z / \partial x - \partial \widetilde{\omega}_x / \partial z
$ that in the physical plane links the perturbation vorticity
components in the $x$ and $z$ directions ($\widetilde{\omega}_x$
and $\widetilde{\omega}_z$) and the perturbed
velocity field ($\widetilde{v}$), $\phi = tan^{-1}(\gamma/\alpha)$ is the perturbation obliquity angle with respect to the $x$-$y$ plane, $k$ is the polar wavenumber, $\alpha = k \cos(\phi)$, $\gamma = k \sin(\phi)$ are the wavenumber components in the $x$ and $z$ directions, respectively, see Fig.1.

Unlike traditional methods where travelling wave normal modes are assumed as solutions, we follow \cite{CJLJ97} and  use arbitrary initial specifications without having to resort to eigenfunction expansions, for more details see Section 2.1. For any initial small-amplitude three-dimensional disturbance, this approach allows the determination of the full temporal behaviour, including both early-time and intermediate transients and the long-time asymptotics. Among all the possible inputs, we focus on arbitrary symmetric and antisymmetric initial conditions distributed over the whole shear region.

\noindent The transversal vorticity $\hat{\omega}_y(y,t)$ is initially taken equal to zero to highlight the three-dimensionality net contribution on its temporal evolution (see \cite{S08,CJLJ97}, to consider the effects of non-zero initial transversal vorticity). Therefore, initial conditions can be shaped in terms of the transversal velocity (see thin curves in Fig. 1b and 1c), as follow:

\begin{eqnarray}
\hspace{-2cm}
&& \,\,\,\,\,\,\,\,\,\,\,\,\, \textmd{Channel flow:} \,\,\,\,\, \hat{v}(y,t=0) = (1-y^2)^2, \,\,\,\,\, \hat{v}(y,t=0) = y(1-y^2)^2, \nonumber \\
\hspace{-2cm}
&& \,\,\,\,\,\,\,\,\,\,\,\,\, \textmd{Wake flow:} \,\,\,\,\, \hat{v}(y,t=0) = \exp(-y^2) \cos(y), \,\,\,\,\, \hat{v}(y,t=0) = \exp(-y^2) \sin(y) \nonumber.
\end{eqnarray}

\noindent For the channel flow no-slip and impermeability boundary conditions are imposed, while for the wake flow uniformity at infinity and finiteness of the energy are imposed.

Equations (\ref{IVP2_fou1})-(\ref{IVP2_fou3}) are numerically solved by the method of lines: the equations are first discretized in the spatial domain using a second-order finite difference scheme, and then integrated in time. For the temporal integration we use an adaptative one-step solver, the Bogacki–-Shampine method \cite{BS1989}, which is an explicit Runge–Kutta method of order three using approximately three function evaluations per step. It has an embedded second-order method which can be used to implement adaptive step size. This method is implemented in the \textsf{ode23} Matlab function \cite{SR1997} and is a good compromise between nonstiff solvers, which give a higher order of accuracy, and stiff solvers, which can in general be more efficient.

\section*{Appendix B. Frequency jumps: details on the phase and velocity field}
\addcontentsline{toc}{section}{Appendix B. Frequency jumps: details on the phase and velocity field}

\subsection*{Details on the velocity, phase and frequency evolution in the surrounding of a frequency jump}

Here, we give further details on the transient evolutions. In particular, we focus on the surroundings of the frequency jumps and on the phase evolution.

\begin{figure}
\begin{center}
\includegraphics[width=0.7\columnwidth]{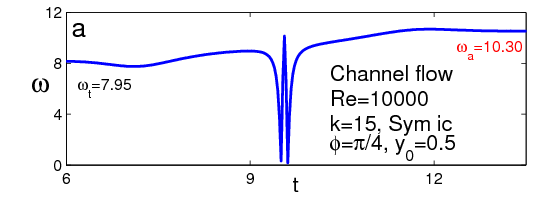}
\end{center}
\begin{center}
\vspace{-0.4cm}
\includegraphics[width=0.7\columnwidth]{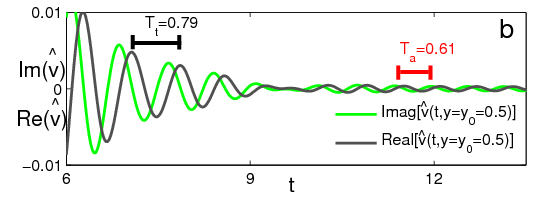}
\end{center}
\begin{center}
\vspace{-0.4cm}
\includegraphics[width=0.7\columnwidth]{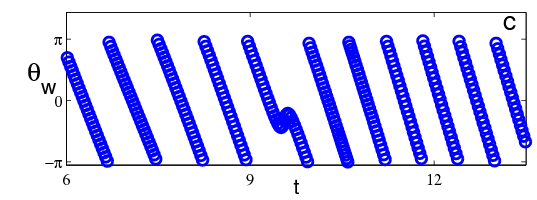}
\end{center}
\begin{center}
\vspace{-0.4cm}
\includegraphics[width=0.7\columnwidth]{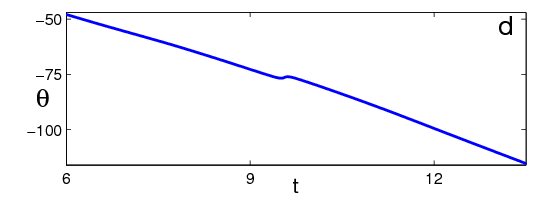}
\end{center}
	\caption{Channel flow, $Re=10000$, $k=15$, symmetric initial condition, $\phi=\pi/4$, observed at $y_0=0.5$, a distance from the wall 1/4 of the channel width. (a) Frequency temporal evolution: $\omega_t$ is the value in the early transient while $\omega_a$ is the asymptotic one. (b) Perturbation transversal velocity $\hat{v}$ (real and imaginary parts). Temporal periods ($T_t$: transient value, $T_a$: asymptotic value). (c) Wrapped wave phase, $\theta_w(t)$. (d) Unwrapped wave phase, $\theta(t)$. For a visualization of the wave solution, see the film \textsf{ChannelFrequency.avi} in the Supplementary Material.}
	\label{imag_real}
\end{figure}

\begin{figure}
\begin{center}
\includegraphics[width=0.7\columnwidth]{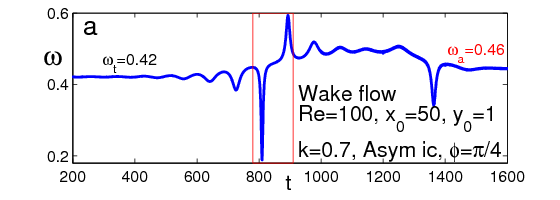}
\end{center}
\begin{center}
\vspace{-0.4cm}
\includegraphics[width=0.7\columnwidth]{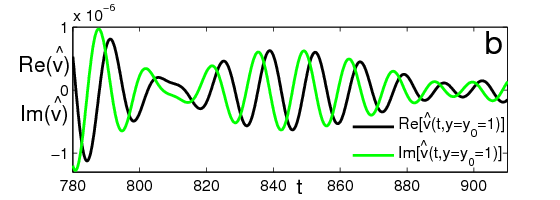}
\end{center}
\begin{center}
\vspace{-0.4cm}
\includegraphics[width=0.7\columnwidth]{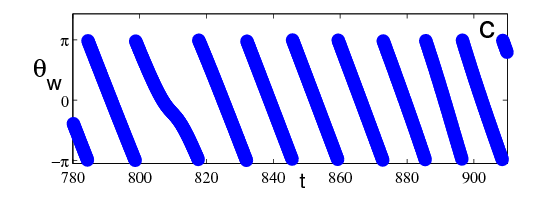}
\end{center}
\begin{center}
\vspace{-0.4cm}
\includegraphics[width=0.7\columnwidth]{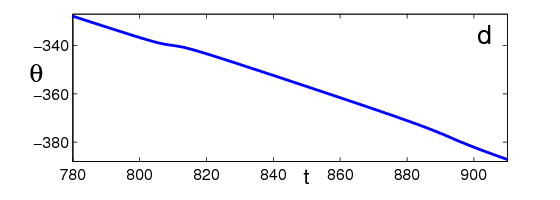}
\end{center}
	\caption{Plane wake flow, $Re=100$, $k=0.7$, symmetric initial condition, $\phi=\pi/4$, observed at $x_0=50$ and $y_0=1$. (a) Temporal frequency evolution: $\omega_t$ is the value in the early transient, $\omega_a$ is the asymptotic value. (b) Perturbation transversal velocity $\hat{v}$ (real and imaginary parts) near the frequency jump highlighted in the panel (a) by the red rectangle. (c)-(d) Wrapped, $\theta_w(t)$, and unwrapped, $\theta(t)$, wave phase in the surroundings of the jump. For a visualization of this wave solution, see the film \textsf{WakeFrequency.avi} in the Supplementary Material.}
	\label{imag_real_wake}
\end{figure}

With reference to the  Fig. \ref{transient}, we show in the following figures enlarged views of the time intervals where jumps occur. The different values of frequency in the early transient ($\omega_t=7.95$) and in the asymptotic state ($\omega_a=10.30$) displayed in Fig. \ref{imag_real}a for the channel flow are caused by an abrupt temporal variation of the phase (see Fig. \ref{imag_real}c-d for the wrapped and unwrapped wave phases temporal evolution). Two distinct temporal periods, $T_t$ and $T_a$, are shown in Fig. \ref{imag_real}b, by highlighting the real (or imaginary) part of the perturbation transversal velocity, $\hat{v}$, at a fixed spatial point, $y_0=0.5$. The discontinuity separates the early time interval where  the transient period is $T_t=0.79$ ($\omega_t=7.95$), from the time interval where the final asymptotic period, $T_a=0.61$ ($\omega_t=10.3$), is reached.

Frequency discontinuity is found for the wake too, see Fig. \ref{imag_real_wake}, where panels b-d refer to the time interval highlighted in red in panel a. The sudden variation of the frequency is due to the phase change, described in panels (c) and (d). Here again, the temporal region where the frequency jumps appear separates the early transient, where the period is $T_t=15$ ($\omega_t=0.42$), from the intermediate transient at the end of which the  asymptotic period, $T_a=13.6$ ($\omega_a=0.46$), is obtained.

\subsection*{Influence of the transversal position where the transient is analyzed}

\begin{figure}
\begin{center}
\begin{minipage}[]{0.49\columnwidth}
\includegraphics[width=\columnwidth]{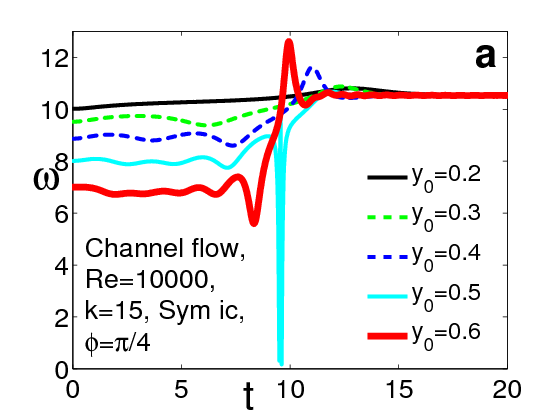}
\end{minipage}
\begin{minipage}[]{0.49\columnwidth}
\includegraphics[width=\columnwidth]{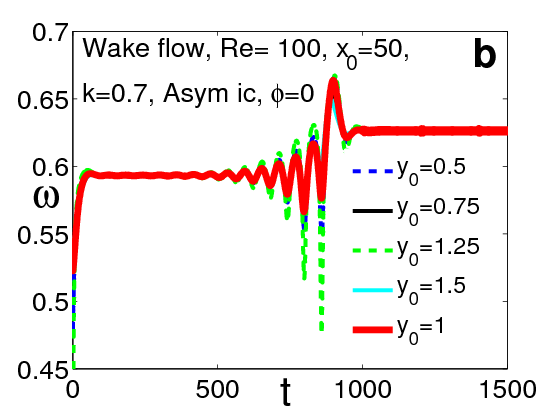}
\end{minipage}
\end{center}
%\vskip -6mm
\caption {Angular frequency, $\omega$, computed at different transversal observation points, $y_0$. (a) Channel
flow, $Re = 10000$, $k = 15$, symmetric initial condition, $\phi = \pi/4$. (b)
Wake flow, $Re = 100$, $k = 0.7$, antisymmetric initial condition, $\phi = 0$, $x_0 = 50$ diameters downstream
the body.}
\label{omega_y0}
\end{figure}

In Fig. \ref{omega_y0}, we show the frequency transient as observed at different transversal points $y_0$, for the channel flow (panel a) and the wake (panel b). The points, $y_0$, are chosen in the high shear region. We consider $y_0$ ranging from $0.2$ to $0.6$ in the channel, and from $0.5$ to $1.5$ in the wake.
By varying $y_0$, the asymptotic values of the frequency remain unaltered, while its transient dynamics can change. This behaviour is more evident for the channel flow case (see panel a). However, for both base flows, the presence of the frequency discontinuity is not affected by the specific choice of $y_0$.

\end{document}